# MIPSGAL 24 μm Observations of Galactic Planetary Nebulae

J.P. Phillips, R.A. Marquez-Lugo

Instituto de Astronomía y Meteorología, Av. Vallarta No. 2602, Col. Arcos Vallarta, C.P. 44130 Guadalajara, Jalisco, México   e-mails : jpp@astro.iam.udg.mx, alejmar.astro@gmail.com,

**Abstract**

We have obtained 24 $\mu$m imaging, profiles and fluxes for 224 planetary nebulae (PNe) lying within the limits of the *Spitzer* MIPSGAL survey. It is noted that most of the PNe having extended 24 $\mu$m emission also possess circular morphologies, suggesting that the emission derives from cool grains located within the AGB mass-loss regimes. Certain of these halos are found to have a surface brightness fall-off which may be consistent with secularly invariant mass-loss within the PNe progenitors. By contrast, the 8.0 $\mu$m envelopes are detected out to smaller distances from the nuclei, and have a steeper rate of surface brightness fall-off; a phenomenon which may arise from changes in the excitation of polycyclic aromatic hydrocarbons (PAHs) within external photo-dissociation regimes (PDRs). Our 24 $\mu$m fluxes are compared to those in previously published studies, and this appears to indicate that many of the prior fluxes have been underestimated; a disparity may imply that previous aperture sizes were too small. We have also combined our 24 $\mu$m fluxes with measures at shorter mid-infrared (MIR) wavelengths, taken with the Infrared Array Camera (IRAC). These are used to investigate the positioning of PNe within the IRAC-MIPSGAL colour planes. The [8.0]-[24] and [5.8]-[24] colours are found to be large, and extend over the respective ranges 3.4-8.7 mag, and 5.4-10.3 mag; indices which are only explainable where a broad range of mechanisms contribute to the fluxes, including PAH bands, cool dust continua, and a variety of ionic transitions. These and other components also affect the morphologies of the sources, and lead to wavelength dependent changes in the widths of the profiles.

**Key Words:** (ISM:) planetary nebulae: general --- ISM: jets and outflows --- infrared: ISM --- (ISM:) dust, extinction

# 1. Introduction

Early ground-based observations of planetary nebulae (PNe) in the near- and far-infrared showed that many of them had excess emission attributable to hot ($\approx 10^3$ K) (see e.g. Willner et al. 1972, 1979; Phillips et al. 1984, 1986; Cohen & Barlow 1974) and cool ($\approx 10^2$ K) (e.g. Forrest et al. 1981; Moseley 1980) dust, although angular resolutions were modest, and the distribution of emission poorly defined. Exceptions in the near infrared (NIR) included A30, NGC 6302 and NGC 2346, which showed evidence for compact emission close to the position of the central stars (e.g. Lester & Dinerstein 1984; Dinerstein & Lester 1984; Cohen & Barlow 1974; Costero et al. 1986), and sources such as IC 418, BD +30° 3639 and NGC 6720 (Bentley et al. 1984; Phillips 1984, 1986; Beckwith et al. 1978), for which emission extends within and outside of the ionized shells.

Subsequent far-infrared (FIR) observations with the Infrared Astronomical Satellite (IRAS) indicated the presence of appreciable masses of cool dust (e.g. Leene, Zhang & Pottasch 1987), and permitted mapping of larger PNe such as NGC 7293 and NGC 6853 (Leene & Pottasch 1987; Zhang et al. 1986). It was apparent that whilst the 60 and 100 $\mu$m emission in NGC 7293 was highly extended, and primarily associated with cool dust continua, the emission at 25 $\mu$m was confined to the centre of the source. This centrally concentrated emission was initially identified as arising from [OIV] $\lambda$25.87 $\mu$m, although *Spitzer* observations suggest a differing interpretation (Su et al. 2007).

More recently, the 2MASS survey has permitted NIR observations for a larger sample of PNe (see e.g. Ramos-Larios & Phillips 2005), whilst the Spitzer Space Telescope (Werner et al. 2004) has been used to obtain mapping in the mid-infrared (MIR), including pointed observations of higher Galactic latitude PNe (e.g. Hora et al. 2004; Su et al. 2007; Guiles et al. 2007; Phillips & Ramos-Larios 2010; Ramos-Larios & Phillips 2009), and larger-scale surveys of lower latitude PNe (e.g. Phillips & Ramos-Larios 2008, 2009; Ramos-Larios & Phillips 2008; Cohen et al. 2007). These latter results involve observations taken with the Infrared Array Camera (IRAC; Fazio et al. 2004), and reveal that the infrared morphologies of the nebulae have a broad variety of forms – structures which are sometimes at variance with those observed in the visible. It is apparent for instance that longer wave emission (at 5.8 and 8.0 $\mu$m) often extends outside of the optically defined envelopes, and may arise from cool dust continua, shock or fluorescently excited $H_2$, and/or bands of polycyclic aromatic hydrocarbons (PAHs) within photo-dissociative regimes (PDRs). In other cases, much of the longer-wave emission appears to arise from close to the nuclei of the sources, and is associated with the central tori of bipolar planetary nebulae (BPNe).

Individually pointed observations at longer IR wavelengths (24, 70 and 160 $\mu$m) have also been taken using the Multiband Imaging Photometer for Spitzer (MIPS; Rieke et al. 2004), and show the presence of extended shells associated with the AGB envelopes (Ueta 2006; Su et al. 2004, 2007). There is also evidence for centrally concentrated 24 $\mu$m emission in a large fraction of PNe, associated with local components of dust (Su et al. 2007) and/or the $\lambda$25.9 $\mu$m transition of [O IV] (Ueta 2006; Su et al. 2004, 2007). Chu et al. (2009) argue that that the extension of 24 $\mu$m emission with respect to the H$\alpha$ shells is informing us of the evolutionary state of the PNe, with the more extended sources corresponding to the earliest phases of evolution.

More recently, the MIPSGAL survey (see http://mipsgal.ipac.caltech.edu/) of the inner 278 deg$^2$ of the Galactic plane has been used to investigate the 24 $\mu$m fluxes and structures of a limited sample of PNe. Thus, Zhang & Kwok (2009) investigated the properties of 37 PNe having IRAC GLIMPSE II counterparts, whence it is found that many of the sources are extended at longer (24 $\mu$m) wavelengths. They also appear to have colour indices and spectra suggesting the presence of PAH emission features. By contrast, Mizuno et al. (2010) have undertaken a more extended survey of 416 extended, resolved, disk- and ring-like objects. They find that ~ 90% of extended objects without central sources are likely to correspond to Galactic PNe.

We have undertaken a more extended MIPSGAL 24 $\mu$m survey for PNe listed by Acker et al. (1992); in the Macquarie/AAO/Strasbourg H$\alpha$ Galactic Catalogue of Galactic Planetary Nebule (MASH I; Parker et al. 2006); and in the MASH II supplement of Miszalski et al. (2008). This results in 24 $\mu$m flux measurements for 224 PNe, which are then used to construct IRAC-MIPSGAL colour diagrams. It is pointed out that the colour indices of these nebulae imply appreciable emission from cool dust continua and PAH emission bands; a result which is in broad agreement with the conclusions of Zhang & Kwok (2009). It is additionally noted that ionic transitions, and various 24 $\mu$m band features may be important in explaining these results.

We shall finally present combined IRAC-MIPSGAL imaging for some 36 PNe, and profiles for 6 of the nebulae having high S/N extended envelopes. These will be used to investigate the morphologies of the sources at differing MIR wavelengths, and determine the radial fall-off in surface brightness. It is pointed out that the varying properties of the emission mechanisms, their differing regimes of excitation, and their differing levels of importance within the photometric channels, probably explains the variations which are observed in profile widths.

## 2. Observations, and Comparison with Previous Results

We have acquired 24 $\mu$m profiles images and fluxes for 224 Galactic PNe. The positions of the sources were taken from the Acker et al. (1992), MASH I (Parker et al. 2006) and MASH II (Miszalski et al. 2008) catalogues, and correspond to the so-called "true" PNe whose identification is regarded as being highly probable. Sources designated as "likely" or "possible" are excluded from this analysis.

The MISPGAL results were taken between 2005 and 2006 using the Multiband Imaging Photometer for Spitzer (MIPS), and produced 24 and 70 $\mu$m imaging of 278 square degrees of the Galactic plane between latitudes -1° < b < 1°, and longitudes l < 62°, l > 298°. We shall be using the publicly available images at 24 $\mu$m taken using a 128x128 Si:As array; mapping which has a sensitivity limit of ~110 mJy, and a saturation limit of 1700 MJy/sr. The fields of view were 5.4x5.4 arcmin$^2$, the resolution was 6 arcsec, and the pixel sizes were 1.25 arcsec. The 24 $\mu$m filter has a 10 % cut-on wavelength of 20.5 $\mu$m, an equivalent cut-off wavelength of 28.5 $\mu$m, and respective average and peak wavelengths of 23.68 $\mu$m and 21.9 $\mu$m. Various distortions are apparent in the imaging results, and these are fully described in the MIPS Instrument Handbook (available at http://ssc.spitzer.caltech.edu/mips/mipsinstrumenthandbook/2/). The most critical artefacts, from the point of view of the present analysis, appear to derive from the optics of the telescope (rather than the camera system). These include radially extended features (or rays) deriving from the mirror support spiders, and rings associated with the telescope diffraction pattern. There is some discussion of the role of these features in Sects. 3 and 4, and care must be taken in distinguishing between diffraction effects and intrinsically extended emission structures. The accuracy of MIPSGAL photometry appears to be reasonably high, and there are few of the systematic errors that were noticed in shorter-wave IRAC results (see below). A discussion of the current understanding of MIPSGAL photometry can be found in Cohen (2009).

We have also combined the MIPSGAL images for certain of the sources with shorter wavelength observations taken by the Galactic Legacy Infrared Mid-Plane Survey Extraordinaire (GLIMPSE). The latter images were taken with IRAC, and employed filters having isophotal wavelengths (and bandwidths $\Delta\lambda$) of 3.550 µm ($\Delta\lambda$ = 0.75 µm), 4.493 µm ($\Delta\lambda$ = 1.9015 µm), 5.731 µm ($\Delta\lambda$ = 1.425 µm) and 7.872 µm ($\Delta\lambda$ = 2.905 µm). The normal spatial resolution for this instrument varies between ~1.7 and ~2 arcsec (Fazio et al. 2004), and is reasonably similar in all of the bands, although there is a stronger diffraction halo at 8 µm than in the other IRAC bands. This leads to differences between the point source functions (PSFs) at ~0.1 peak flux.

Profiles were also evaluated for six of the sources in the IRAC and 24 µm bands, a process which involved the subtraction of background sky fluxes (particularly strong at 5.8-24 µm), and smaller lineal gradients in background emission. Some care needs to be taken in the case of the relative IRAC results, since scattering within the camera is known cause systematic errors in photometry. The flux corrections described in Sect. 4 of the IRAC Instrument Handbook (Carey et al. 2010) suggest that surface brightness estimates need to be corrected in extended sources, and that maximum changes are of the order of ~0.91 at 3.6 µm, 0.94 at 4.5 µm, 0.66-0.73 at 5.8 µm, and 0.74 at 8.0 µm; although the precise values of these corrections also depend on the distribution of source emission. Such changes do not affect present profiles to any appreciable degree, however, and make no difference at all to our analysis or conclusions.

We have finally obtained fluxes for all PNe for which there is evidence of emission in the 24 μm photometric channel. Levels of background are assessed at six hexagonally located positions about the nebulae, selected so as to avoid extraneous emission from unrelated field stars, or from the outer halos of the nebulae themselves. Taken as a whole, and deriving an average at high and low latitudes, we determine that there are ~ 1.3 stars per arcmin$^2$ capable of appreciably affecting the current photometry. Given the mean beam size quoted below, this implies that ~13 % of fluxes have a possibility of being contaminated by such stars. Where field stars lie close to the central nebulae, on the hand, we have attempted to adjust the aperture sizes so as to avoid contamination. This leaves the actual total of sources for which contamination proved critical at just ~ 8 % of the total in Table 1 (17 PNe in all). These contaminants were separately assessed, and eliminated from the results.

Although contamination by brighter field stars is therefore expected to be small, it is possible that certain of the results are affected by fainter stellar components – stars which are very much weaker, and difficult to distinguish from the nebular emission. Whilst these may increase certain of the fluxes, their contribution is expected to be small, and they would have little influence on the analysis below.

The sources are measured irrespective of whether there is evidence for shorter wave IRAC emission or not, and is limited to nebulae which are clearly centred within the photometric apertures employed. Any sources which are located outside of these apertures, whether extended or not, have been excluded from the present analysis. We note that several of the nebulae measured by Zhang & Kwok (2009) and Mizuno et al. (2010) are not detected in these apertures, or lie sufficiently outside of them that their identification is in doubt.

A further critical point in determining the present fluxes is that the aperture sizes should be sufficiently large, and able to capture the majority of PN flux. Given that certain sources appear to have extended halos, and that the halos are not always apparent from a cursory inspection of the images, it is necessary to employ apertures which are significantly larger than the bright central cores, but not so large as to lead to problems with field star/background contamination.

A comparison of our aperture sizes B for 25 nebulae common to the list of Mizuno et al. (2010), and for which Mizuno et al. provide nebular sizes $\theta_{PN}$, suggests mean values $<B/\theta_{PN}>$ = 2.2, with a minimum value $B/\theta_{PN}$ = 1.44.

It is pertinent, in this respect, to determine how close present fluxes agree with previous photometry. In the study of Mizuno et al. (2010), for instance, the authors state that the radii of the apertures is selected so as to "minimally contain the entire object". Whilst the precise meaning of the phrase is open to some question, it is clear that many of their measured fluxes are less than is observed here. A cross correlation of the results is illustrated in Fig. 1, for instance. Whilst the trend of points gives the impression of a linear one-to-one agreement, with the source having the highest flux (IC 4673) appearing to be the most discrepant, it is clear that the one-to-one variation, and least-squares trends are very slightly at variance with one another. A more detailed analysis shows greater cause for concern. Some 30% of the sources have 24 $\mu$m fluxes which differ by > 15% from those which are listed here, with the most extreme cases (PHR1753-2905 and BMP1749-2356) differing by 121 and 410%. All but one of these residuals implies F(Mizuno et al. ) < F(Present), suggesting that many of the Mizuno et al. results may seriously underestimate the source fluxes. Where these disparate sources are included, then the mean value <F(Present)/F(Mizuno et al.)> = 1.065. Where they are excluded, on the other hand, then the ratio becomes 1.003. It therefore follows that where the most disparate results are deleted from the analysis, then agreement with the Mizuno et al. photometry is very good indeed.

A comparison with the results of Zhang & Kwok (2009) is also shown in Fig. 2, where it would appear that there is again a difference between the least-squares and one-to-one trends. For this case, as many as half of the PNe have flux discrepancies > 15 %, with all of them implying F(Present)/F(Zhang & Kwok) > 1. The mean ratio for all of the sources is <F(Present)/F(Zhang & Kwok)> = 1.06, which declines to 1.03 when the discrepant sources are removed.

It is therefore clear that selection of aperture size is very important indeed, and it is likely that many of the discrepancies cited above arise from the use of too small an aperture dimension.

The present PNe fluxes are listed in Table 1, where we provide the Galactic positional identification (column 1); the alternative source name (column 2); the catalogues in which the PNe are listed (column 3); and the right ascensions and declinations (columns 4 & 5). The aperture diameter is indicated in column 6, and the fluxes in mJy are provided in column 7. The total errors in the fluxes depend partially upon the Poisson noise; although by far the most serious errors are normally associated with variations in the background emission (see e.g. the discussion in Mizuno et al. (2010), as well as the MIPS Instrument Handbook). We provide an estimate of this latter uncertainty in column 8 based on variations in the six background apertures about each of the sources. Finally, we have used the zero magnitude calibration of Engelbracht et al. (2007) to determine magnitudes [24], and this and the mean error in [24] (denoted $\sigma([24])$) are listed in the final two columns of Table 1.

## 3. MIPSGAL and GLIMPSE imaging of Galactic Planetary Nebulae

MIPSGAL imaging of 4 PNe (PHR 1743-2431 & 1722-3210, and MPA 1748-2402 & 1739-2702), and combined IRAC and MIPSGAL imaging of a further 32 PNe is illustrated in Fig. 2, where several aspects of the sources are immediately apparent:

(i) Almost all of the nebulae have 24 $\mu$m envelopes which are larger than those of the corresponding shorter-wave (IRAC) structures. The possible exceptions include A 48, HaTr 10, A 53, and PHR 1552-5254. The structures in all of these latter cases appear to possess a flocculent appearance which is likely to be real (i.e. not the result of noise in the results).

(ii) A good fraction of all of the sources (~ 40%) have very weak IRAC emission components, and their images are dominated by the 24 $\mu$m envelope alone.

(iii) Almost all of the 24 $\mu$m envelopes have a circular morphology, particularly where levels of IRAC emission are low. The only exceptions appear to be PHR 1246-6324 and He 2-111, the IRAC images for which have been considered by Phillips & Ramos-Larios (2008) and Ramos-Larios & Phillips (2008). Both of the sources have shorter-wave ring-like structures, probably (in the case of He 2-111) related to an exterior bipolar outflow. The orientation of the 24 μm shell about PHR 1246-6324 is similar to that of the inner nebular ring – both have similar major axis position angles.

Two further sources (MPA 1840-0529 & 1739-2648) have extended envelopes, but have been excluded from Fig. 2. It is likely that the larger part of their extended emission is attributable to diffraction. This may also apply, to some degree, for the sources BMP 1524-5746, MPA 1747-2649, and MPA 1523-5710, for which there is evidence of dark annular rings corresponding to the first Airy minimum.

It is therefore apparent that ~ 85 % of extended sources have envelopes which are usually larger than is discerned at shorter IRAC wavelengths. It is also likely, given the apparent morphologies and sizes of these structures, that we are observing cool dust emission and/or PAH emission bands within the AGB mass-loss halos (see also our further comments in Sect. 4). The remaining (large majority of) sources are presumably physically more compact, and/or located at greater distances from the Sun. Several may also correspond to older PNe (Chu et al. 2009), with centrally concentrated 24 μm emission.

## 4. MIPSGAL and IRAC Profiles

In addition to the imaging described in Sect. 3, we have determined IRAC and 24 μm profiles for several PNe. The results are shown in Fig. 3.

A superficial inspection of the profiles gives the impression that source sizes increase with increasing MIR wavelength. The actual variation, however, turns out to be less simple or so extreme. Where the deconvolved FWHM of the profiles is given by $\theta_{PN} \cong (\theta_{OBS}^2 - \theta_{PSF}^2)^{0.5}$, for instance, and one defines a normalised dimension through $\theta_{PNN} = \theta_{PN}[\lambda]/\theta_{PN}[8.0]$, then we obtain the results indicated in Table 2; where $\theta_{OBS}$ is the observed FWHM of the source, $\theta_{PSF}$ is the width of the PSF, and we have indicated dimensions for eight PNe (including the six in Fig. 3). In certain cases, the 3.6 μm profiles were strongly affected by field star emission, and we have excluded these results from the data in Table 2. Similarly, one source (PPA 1755-2739) was unresolved at 24 μm, and the dimensions of the nebula are poorly determined. Finally, it is worth noting that this deconvolution technique is only applicable for Gaussian emission fall-offs – although it leads to little uncertainty where values of $\theta_{PN}[8.0]$ and $\theta_{PN}[24]$ are large, and should provide a reasonable guide to variations in $\theta_{PNN}$.

It would seem, from looking at the runs of estimates for $\theta_{PNN}$, that the values of $\theta_{PNN}[5.8]$ and $\theta_{PNN}[24]$ are mostly greater than unity – that is, the 5.8 and 24 μm profile widths are greater than those at 8.0 μm. In neither case, however, is the difference more than ~60%, and it is for the most part very much less. There are also several cases where $\theta_{PNN}[24]$ is much less than unity, from which it would appear that 24 μm emission is concentrated in the centres of the sources.

Finally, it is clear that the 3.6 and 4.5 μm source widths are mostly smaller than $\theta_{PN}[8.0]$, and that $\theta_{PNN}[4.5] > \theta_{PNN}[3.6]$.

It is therefore apparent that variations in FWHM are relatively modest, although systematic variations in these parameters are nevertheless apparent. Such variations can be largely understood through the roles of differing emission mechanisms, and their relative importance in the differing photometric channels (see Sect. 5). Thus, where ionised emission is dominant in the 3.6 μm channel, then one might expect that the dimensions of the sources would be comparable to those in the visual, where emission is dominated by permitted and forbidden line transitions. As one passes towards 5.8 μm, on the other hand, where ionic line emission is weak, but 6.2 μm PAH band emission is likely to be strong, then the location of PAH band carriers in exterior PDRs might be expected to lead to the broadening of the profiles evident in Table 2.

The 8.0 μm band is dominated by transitions such as [ArII] $\lambda$6.985 μm, [ArIII] $\lambda$8.991 μm, [Ne VI] $\lambda$7.642 μm and [Ar V] $\lambda$7.901 μm, as well as strong PAH emission bands at 7.7 and 8.6 μm. The role of the ionic transitions may explain why $\theta_{PN}[8.0]$ is so very often less than $\theta_{PN}[5.8]$. Finally, it is apparent that both cool dust emission, and transitions such as [Ne V] $\lambda$24.32 and [O IV] $\lambda$25.87 μm may play a role in differing sectors of the nebular shells – and explain why some nebulae possess values $\theta_{PNN}[24] > 1$, and others are more centrally concentrated ($\theta_{PNN}[24] << 1$).

Whilst the IRAC profiles of the sources show clear double-peaked structures, this morphology is very much softened in the longer-wave 24 μm channel. There are various factors which may contribute to this trend. The PSF at 24 μm is significantly larger than for the shorter wave results, for instance, and we have indicated this function in the plots in Fig. 3. It is clear that this is, on its own, capable of explaining a larger part of the infilling observed here. It would not however account for the trends in Th 2-A and PHR 1619-4914. It is also clear that substantial halo components of cool dust emission will also soften the 24 μm profiles (see our further discussion below), and decrease the prominence of interior shell-like structures, whilst a good fraction of nuclear 24 μm emission may also be associated with the $\lambda 24.32$ and $\lambda 25.87$ μm transitions of [Ne V] and [O IV]. These are likely to be stronger at smaller radii, and would lead to a reduction in the double peaked morphology.

We have finally illustrated the logarithmic variation of longer wave emission in Fig. 4, where we show profiles for three of the nebulae, together with a 24 μm PSF. Several conclusions may be drawn from a cursory inspection of these results.

The first is that the PSF has a complex surface-brightness fall-off, and possesses several features arising from the telescope optics. Certain of these features are labelled in the lower part of the graph. Excepting the case of NGC 4673, however, where there is evidence for a deviation in surface brightnesses close to r ~ 30 arcsec, such artefacts appear to have little influence on our 24 μm results. All of the profiles fall-off in a smooth and continuous manner out to distances of order r ~ 35 arcsec. There is evidence, beyond this point, for slight changes in gradient which may be related to intrinsic changes in the halos, or to errors in background subtraction. The fall-off over the range for which such gradients can be reliably determined (say r ~10-35 arcsec) turns out, however, to be reasonably consistent. Least-squares fits to the surface brightness fall-offs implies that they can be well represented by a variation ~ $r^{-\alpha}$, with $\alpha$ taking the values ~3.1 for He 2-111, and ~3.55 for Th 2-A and NGC 4673. These fits are indicated in Fig. 4 using grey dashed lines. Typical errors in these exponents appear to be of order ~0.2, whilst the value of α for the PSF is ~2.4. Given that the halos have spherical morphologies; that grain temperatures are reasonably invariant; and that the volume densities of the grains are proportional to the mass density of the associated gas, then it would seem that the emission fall-offs are consistent with a ≈ $r^{-2.4}$ decline in gas density. Where the expansion velocities of the halos are also invariant with radius (and there is some evidence for this in nebulae such as NGC 6543, NGC 6751 and NGC 40 (Chu et al. 1991; Bryce et al. 1992; Meaburn et al. 1996)), then this would suggest that AGB mass loss rates were secularly invariant, or varied relatively slowly over the period of mass loss.

By contrast, we have also indicated the decline in surface brightnesses for the 8.0 μm band – a waveband which is dominated in many cases by PAH emission bands (although ionic transitions and dust continua also contribute to such fluxes). We have traced these profiles out to as far as is reliably possible, whence it is clear that the observed 8.0 μm envelopes are significantly smaller than those at 24 μm. This is hardly surprising given the high surface brightnesses in the 24 μm channel, and the limiting sensitivities of the IRAC results. It does not imply that the 24 μm halos are necessarily > 3 times greater than at shorter wavelengths. However, there does appear to be some evidence that 8.0 μm emission falls-off more steeply, and that exponents α may approach ~4.5. A similar tendency (for steep 8.0 μm fall-offs) has also been noted by Phillips & Ramos-Larios (2010), Ramos-Larios et al. (2008) and Phillips et al. (2009), and may reflect strongly varying levels of excitation for the PAH emission bands. A final caveat to these results concerns the role of scattering within IRAC, and it is possible that at least some of the extended emission may be non-intrinsic to the sources. Although we believe that this effect is small for the cases in Fig. 4, it is possible that it affects the surface brightness fall-offs, and extends the range of 8.0 μm emission.

It is finally worth commenting upon several of the more specific properties of the nebulae in Fig. 3. Th 2-A has a closely circular structure at IRAC wavelengths which has previously been investigated by Phillips & Ramos-Larios (2008). The Hα and 8.0 μm emission extends over ~30 arcsec (see also Gorny et al. 1999), a dimension which is significantly smaller than is observed at 24 $\mu$m (~80 arcsec; see Fig. 4). The IRAC structure of He 2-111, by contrast, takes the form of a narrow high ellipticity ring which is oriented perpendicularly to the optical bipolar structure (see e.g. Corradi & Schwarz (1995) for an optical image of the source, and Phillips & Ramos-Larios (2008) for a discussion of IRAC mapping results). There is some evidence that the 24 $\mu$m envelope is also elongated in the same direction as the bipolarity, although it appears to possess an ellipsoidal morphology, and can only be traced out to r ~ 70 arcsec; a size which is ≈ 5 times smaller than that of the optical bipolar lobes.

PHR1619-4914 (G333.9+00.6) is described by Ramos-Larios & Phillips (2008) as being "something of beauty", and consists of a circular shell with well-defined rim centred upon a bright central star (or unresolved nuclear emission envelope). The H$\alpha$ dimensions given by Parker et al. (2006) (35x32 arcsec$^2$) are comparable to those at 8.0 $\mu$m, although there is also evidence for an MIR halo extending over ~ 80 arcsec or so. The 24 $\mu$m envelope, by contrast, appears to have dimensions of ~ 130 arcsec.

Finally, we have included PPA 1755-2739 as an example of a more compact emission source for which the 24 $\mu$m shell is barely resolved – the profile is only ~ 23 % broader than that of the relevant PSF. This centrally concentrated emission is also associated with a broader envelope which appears to extend to relative positions (RPs) of ±15 arcsec. A comparison of this latter variation with that of the PSF, however, reveals that much of the emission is attributable to diffraction – an example of where care needs to be taken in the interpretation of the results. It is also noteworthy that the IRAC profiles are markedly different from each other; there appears to be a right-to-left shift in emission peaks as one passes from shorter to longer MIR wavelengths. Although this may represent an intrinsic change in the properties of the source, perhaps not dissimilar to those noted previously for G302.3-00.5 (Ramos-Larios & Phillips 2008; also included as source PHR1246-6324 in Fig. 2), it is also possible that we are observing field star contamination at the shorter IRAC wavelengths; a factor which may also affect overall fluxes, and cause a reduction in indices [3.6]-[4.5].

It is finally apparent that 24 μm fluxes are very much larger than those in the shorter-wave IRAC channels. This leads to large IRAC-MIPSGAL colour indices, as will be discussed in Sect. 5, where we consider the various emission mechanisms in somewhat more detail.

## 5. Colour-Colour Diagrams and the Roles of Differing Emission Mechanisms

We have combined the 24 μm photometric results from Table 1 with IRAC photometry deriving from Phillips & Ramos-Larios (2008, 2009) & Ramos-Larios & Phillips (2008) to create the [3.6]-[5.8]/[8.0]-[24] colour-colour map illustrated in Fig. 5. The Acker et al. (1992) sources are represented as grey diamonds, MASH I nebulae by open circles, and MASH II PNe by solid triangles. We have also included the trends to be expected for a blackbody (BB) dust continuum, and combined BB + PAH emission components, whilst indicating the vectors associated with ionized emission, $H_2$ transitions, and various 24 μm band features.

It is apparent that the [8.0]-[24] indices are appreciable, and extend over the range 3.4-8.7 mag. They are also well outside of the range expected for BB emitting grains. Where dust emission coefficients vary as $\sim \lambda^{-\gamma}$, on the other hand, and $\gamma > 0$, then the relevant dust continuum loci would move even further down and to the lower right. We shall therefore consider BB dust continua alone in the following analysis, whilst noting that larger values of $\gamma$ lead to similar results.

The role of ionized emission depends strongly upon the nature of the central stars, and the excitation level of the PNe. An analysis by Reach et al. (2006), however, suggests that ionized gas would normally lead to fluxes 0.25/3.7/0/1 in the B1/B2/B3/B4 IRAC bands. Given that values such as these apply for most PNe, then one expects that [3.6]-[5.8] will decrease as levels of ionic line emission become larger. The 8.0 μm band also has appreciable levels of emission arising from the [ArII] $\lambda$6.985 μm and [ArIII] $\lambda$8.991 μm lines, or the higher excitation [Ne VI] $\lambda$7.642 μm and [Ar V] $\lambda$7.901 μm transitions, and this would tend to lead to smaller values of [8.0]-[24] in the absence of appreciable emission from [O IV] $\lambda$25.87. On the other hand, it has been argued that this latter transition may be strong in several PNe (Ueta 2006; Su et al. 2004, 2007), and would lead to enhanced levels of flux at smaller distances from the central stars. Where this is the case, then [8.0]-[24] indices may increase.

The 24 $\mu$m fluxes of PNe are strongly influenced by cool dust emission, and this undoubtedly represents a crucial factor in understanding the present results. It is also worth noting however that the 24 $\mu$m passband, which extends between ~ 20.5 $\mu$m and 28.5 $\mu$m (see Sect. 2), is affected by several further components of emission. These include the so-called 21 $\mu$m feature (between ~ 19 and 24 $\mu$m), and the 30 $\mu$m feature (between ~ 27 and 42 $\mu$m). The origins of the former component are still far from clear, and of order ~ 10 or so mechanisms have been suggested to explain this excess. Most recently, Zhang, Jiang & Li (2009) have suggested nano-FeO as the most likely carrier of these bands. The main carrier for the 30 $\mu$m feature is also far from clear, although it may correspond to MgS, or some other still unidentified carbonaceous material (see e.g. Hony et al. 2002). Both of these features appear to be associated with carbon-rich PNe, although the 21 $\mu$m feature has mostly been detected in younger outflow sources.

Apart from this, oxygen rich sources also have their complement of emission bands, mostly associated (at these wavelengths) with crystalline silicate transitions (see e.g. Molster & Kemper 2005). It is therefore clear that all manner of bands may contribute to the fluxes, and drive [8.0]-[24] to higher positions in Fig. 5.

$H_2$ emission has also been observed in many PNe, and may be excited in post-shock regimes, or by fluorescent excitation by the central star (e.g. Graham et al. 1993; Guerrero et al. 2000; Kastner et al. 1996; Ramos-Larios et al. 2006, and references therein). The shocked components of $H_2$ are expected to lead to relative intensities 0.42/0.52/0.90/1 in the respective B1-B4 IRAC bands (Reach et al. 2006). If these values are also taken as being typical of PNe, then it follows that [3.6]-[5.8] indices might be expected to increase. By contrast, the 24 $\mu$m channel contains the S(0) v = 0-0 $\lambda$28.22 $\mu$m transition of $H_2$, whilst the 8 $\mu$m channel has the S(4) $\lambda$8.02 and S(5) $\lambda$6.91 v=0-0 lines. It seems likely that [8.0]-[24] indices will tend to decrease.

The total effect of these mechanisms is therefore somewhat unclear, and much depends upon the relative strengths of the differing emission mechanisms.

The influence of extinction $A_V$ is also less than well defined. Indebetouw et al. (2005) find that $A_{3.6}$-$A_{5.8} \cong 0.13 A_K$, for instance, although values of $A_\lambda$ for $\lambda$ = 4.5-8.0 $\mu$m are generally the same. On the other hand, Phillips & Marquez-Lugo (2010) have found that the 3.6-to-24 $\mu$m values of $A_\lambda$ may fall-off steeply in several clouds, and in a manner which is similar to that expected for standard Galactic extinction curves. We have therefore used the trends of Cardelli et al. (1989) to determine the reddening vector in Fig. 5, and this is likely to be at least qualitatively correct for most plausible reddening curves. The critical message appears to be that whilst reddening would be capable of explaining the present results, it would require levels of extinction which are implausibly large.

It finally falls to us to mention one further component of emission which may help explain our present results: the 3.3, 6.2, 7.7 and 8.6 $\mu$m PAH band features. Reach et al. (2006) estimate that these have relative intensities 0.054/0.40/1 in the B1/B3/B4 channels. Where we combine these components with dust continuum emission, then one obtains the various dashed curves to the right of Fig. 5. The individual ticks correspond to differing values of grain temperature $T_{GR}$, with each tick separated by $\Delta T_{GR}$ = 10K. The curves are also parameterised in terms of a coefficient $\Lambda = I_{3.6}(PAH+DUST)/I_{3.6}(DUST)$, where $I_{3.6}(DUST)$ is the strength of the dust continuum within the 3.6 $\mu$m photometric channel, and $I_{3.6}(PAH+DUST)$ is the strength of PAH and dust emission combined. The values of $\Lambda$ are indicated in the rectangular boxes. Given that 3.6 $\mu$m channel lies mostly at the Wien limits of the dust continuum trends, it follows that values of $\Lambda$ may be extremely high.

It is apparent that the combined dust + PAH emission does a very good job of increasing indices [8.0]-[24], but that the values are stubbornly fixed to the right-hand side of the graph. It therefore follows that these two components alone cannot explain our present results. However, when one combines PAHs and dust with the various ionic transitions, then one can expect both the high values of [8.0]-[24], and the much lower indices [3.6]-[5.8]. it is therefore clear that our results are only explainable where various emission mechanisms play a role.

It is finally interesting to undertake a similar analysis for the [5.8]-[24]/[3.6]-[4.5] colour plane, illustrated in Fig. 6. In this case, the combined BB continuum + PAH emission bands have no trouble at all in explaining the present results, although the dust temperatures of many of the sources are required to be rather high - much larger than the ≈60-150K values normally associated with these sources (e.g. Stasinska & Szczerba 1999). This disparity might be explained where other components ($H_2$ emission, and short- and longer-wave ionic transitions) drive PNe downwards and to the right.

It is therefore clear that the profiles, imaging and fluxes of the PNe permit us to gain an overall perspective concerning the role of differing emission mechanisms, and their relative importance in defining the properties of the PNe.

## 6. Conclusions

We have obtained fluxes, profiles and images for PNe lying within the angular limits of the 24 $\mu$m MIPSGAL survey, and compared the results to data acquired from the GLIMPSE MIR surveys. It is found that some of the sources (~ 15%) have broadly extended structures which are likely to arise from dust emission within AGB mass-loss regimes. These sources are, for the most part, circularly symmetric, and have a surface brightness fall-off $\propto r^{-\alpha}$, where $3.1 < \alpha < 3.55$. Where grain temperatures are broadly invariant; grain number densities are proportional to the mass densities of halo gas; and finally, where gaseous outflow velocities are invariant with radial distance from the central stars, then such results would imply progenitor mass-loss rates which are secularly invariant, or which vary very slowly over the period of mass loss. By contrast, the surface brightness of 8.0 $\mu$m emission is observed to fall-off more steeply, and also extends over a more limited range of distances from the core, whilst the 3.6-4.5 $\mu$m fluxes are very much more compact, and associated with the nebular ionized cores. Such tendencies also lead to changes in the widths $\theta_{PN}$ of the emission profiles, such that $\theta_{PN}$ tends to be larger at longer wavelengths, and smaller at 3.6 and 4.6 $\mu$m.

Such trends would be understandable where 3.6 and 4.5 μm emission is dominated by ionised transitions (and/or shock or fluorescently excited transitions of $H_2$); the 5.8 emission is strongly influenced by PAH emission bands; 8.0 μm profiles are influenced by ionic transitions and PAH emission bands, and where extended 24 μm fluxes are affected by cool dust continua. Certain of the cases in which $\theta_{PN}[24]$ is small may correspond to sources in which the [Ne V] $\lambda 24.32$ and/or [O IV] $\lambda 25.87$ μm transitions are appreciable, and enhance emission within the cores.

We have also presented 24 μm fluxes for 224 PNe, using aperture sizes which are ~ twice as large as those of the primary ionised shells. This ensures that most of the weaker, more extended emission is included in the results, and leads to fluxes which are significantly greater than those of two previous analyses. These results are combined with prior IRAC photometry to undertake an analysis of the positions of these sources within the MIPSGAL-IRAC colour planes. It is found that [8.0]-[24] and [5.8]-[24] indices are relatively large, and cannot be understood in terms of any single emission mechanism. Fluxes are likely to be dominated by several differing components, among which ionic transitions, dust continua, $H_2$ transitions, and PAH and 30 μm emission bands are likely to be among the most critical for understanding these present trends.

**Acknowledgements**

This work is based, in part, on observations made with the Spitzer Space Telescope, which is operated by the Jet Propulsion Laboratory, California Institute of Technology under a contract with NASA.19

**References**

Acker A., Ochsenbein F., Stenholm B., Tylenda R., Marcout J., Schohn C., 1992, Strasbourg-ESO Catalogue of Planetary Nebulae, ESO, Garching

Beckwith S., Gatley I., Persson S. E., 1978, ApJ, 219, 33

Bentley A.F., Hackwell J.A., Grasdalen G.L., Gehrz R.D., 1984, ApJ, 278, 665

Bryce M., Meaburn J., Walsh J.R., Clegg R.E.S., 1992, MNRAS, 254, 477

Cardelli J.A., Clayton G.C., Mathis J.S., 1989, ApJ, 345, 245

Carey S., Surace J., Glaccum W., Lowrance O., Lacy M., Reach W., 2010, IRAC Instrument Handbook (http://ssc.spitzer.caltech.edu/irac/iracinstrumenthandbook/home/

Chu Y.-H., Manchado A., Jacoby G.H., Kwitter K.B., 1991, ApJ, 376, 150

Chu Y.-H., et al., 2009, AJ, 138, 691

Cohen M., 2009, AJ, 137, 3449

Cohen M., Barlow M.J., 1974, ApJ, 193, 401

Cohen M., et al., 2007, ApJ, 669, 343

Corradi R.L.M., Schwarz H.E., 1995, A&A, 293, 871

Costero R., Tapia M., Mendez R.H., Echevarria J., Roth M., Quintero A., Barral J.F., 1986, Rev. Mex. Astron. Astrofís.. 13, 149

Dinerstein H.L., Lester D.F., 1984, ApJ, 281, 702

Engelbracht C.W., et al., 2007, PASP, 119, 994

Fazio, G., et al. , 2004, ApJS, 154, 10

Forrest W.J., Houck J.R., McCarthy J.F., 1981, ApJ, 248, 195

Gorny S.K., Schawartz H.E., Corradi R.L.M., van Winckel H. Van, 1999, A&ASS, 136, 145

Graham J.R., Herbst T.M., Matthews K., Neugebauer G., Soifer B.T., Serabyn E., Beckwith S., 1993, ApJ, 408, L105

Guerrero M.A., Villaver E.D., Manchado A., Garcia-Lario P., Prada F., 2000, ApJS, 127, 125

Guiles S., Bernard-Salas J., Pottasch S.R., Reillig T.L., 2007, ApJ 2007, 1282

Hony S., Waters L.B.F.M., Tielens A.G.G.M., 2002, A&A, 390, 533

Hora J.L., Latter W.B., 1994, ApJ, 437, 281

Hora J.L., Latter W.B., Allen L.E., Marengo M., Deutsch L.K., Pipher J.L., 2004, ApJS, 154, 296

Hora J.L., Latter W.B., Smith H.A., Marengo M., 2006, ApJ, 652, 426

Indebetouw R., et al., 2005, ApJ, 619, 931

Kastner J.H., Weintraub D.A., Gatley I., Merrill K.M., Probst R.G., 1996, ApJ, 462, 777

Leene A., Zhang C.Y., Pottasch S.R., 1987, in A. Preite-Martinez, ed, Proceedings of the Frascati Workshop, Planetary and Proto-Planetary Nebulae: From IRAS to ISO. D. Reidel Publishing Co., Dordrecht, Holland, p. 39

Lester D.F., Dinerstein H.L., 1984, ApJ, 281, 67

Meaburn J., Lopez J. A., Bryce M., Mellema G., 1996, A&A, 307, 579

Miszalski B., Parker Q.A., Acker A., Birkby J.L., Frew D.J., Kovacevic A., 2008, MNRAS, 384, 525

Mizuno D.R., et al. 2010, AJ, 139, 1542

Molster F., Kemper C., 2005, Space Science Reviews, 119, 3

Mosely H., 1980, ApJ, 238, 892

Parker Q. A., Acker A., Frew D. J., Hartley M., Peyaud A. E. J., Ochsenbein F., Phillipps S., Russeil D., Beaulieu S. F., Cohen M., Köppen J., Miszalski B., Morgan D. H., Morris R. A. H., Pierce M. J., Vaughan A. E. , 2006, MNRAS, 373, 79

Phillips J. P., Mampaso A., Vilchez J. M., Gomez P., 1986, Ap&SS, 122, 81

Phillips J.P., Marquez-Lugo A., 2010, MNRAS, in press

Phillips J.P., Ramos-Larios G., 2008, MNRAS, 383, 1029

Phillips J.P., Ramos-Larios G., 2009, MNRAS, 396, 1915

Phillips J.P., Ramos-Larios G., 2010, MNRAS, in press

Phillips J.P., Ramos-Larios G., Schroeder G., Contreras-Verbena J. L., 2009, MNRAS, 399, 1126

Phillips J.P., Sanchez-Magro C., Martinez Roger C., 1984, A&A, 133, 395

Ramos-Larios, G., Phillips, J.P., 2005, MNRAS, 357, 732

Ramos-Larios G., Phillips J.P., 2008, MNRAS, 390, 1014

Ramos-Larios G., Phillips J.P., 2009, MNRAS, 400, 575

Ramos-Larios G., Phillips J.P., Cuesta L., 2008, MNRAS, 391, 52

Ramos-Larios G., Kemp S.N., Phillips J. P, 2006, RMxAA ,42, 131

Reach W.T., et al., 2006, AJ, 131, 1479

Rieke G., et al. 2004, ApJS, 154, 25

Stasinska G., Szczerba R., 1999, A&A, 352, 297

Su K.Y.L., Chu Y.-H., Rieke G. H., Huggins P.J., Gruendl R., Napiwotzki R., Rauch T., Latter W. B., Volk K., 2007, ApJ, 657, 41

Su K.Y.L., et al., 2004, ApJS, 154, 302

Werner M., et al., 2004, ApJS, 154, 1

Willner S.P., Becklin E.E., Visvanathan N., 1972, ApJ, 175, 699

Willner S.P., Jones B., Puetter R.C., Russell R.W., Soifer B.T., 1979, ApJ, 234, 496

Zhang C. Y., Mo J. E., Leene A., Pottasch S. R., 1987, A&A, 178, 247

Zhang K., Jiang B. W., Li A., 2009, MNRAS, 396, 1247

Zhang Y., Kwok S., 2009, ApJ, 706, 252

TABLE 1

24 μm Photometry of MIPSGAL Planetary Nebulae

| PNG | NAME | CAT. | R.A. (2000) H:M:S | DEC (2000) D:M:S | B arcsec | FLUX(24) mJy | $\sigma(24)$ mJy | [24] mag | $\sigma([24])$ mag |
|---|---|---|---|---|---|---|---|---|---|
| G000.0-02.1 | MPA1754-2957 | MASH2 | 17 54 04.3 | -29 57 27 | 48 | 3.49E+02 | 9.13E+00 | 3.28 | 0.03 |
| G000.1+02.6 | Al 2-J | ACKER | 17 35 35.38 | -27 24 03.4 | 30 | 5.14E+02 | 2.61E+00 | 2.86 | 0.01 |
| G000.1-01.1 | M 3-43 | ACKER | 17 50 24.21 | -29 25 17.9 | 48 | 3.27E+03 | 3.14E+01 | 0.85 | 0.01 |
| G000.1-01.7 | PHR1752-2941 | MASH1 | 17 52 48.9 | -29 41 59 | 50 | 6.82E+02 | 3.50E+00 | 2.55 | 0.01 |
| G000.2-01.9 | M 2-19 | ACKER | 17 53 45.91 | -29 43 46.3 | 30 | 7.92E+02 | 2.10E+00 | 2.39 | 0.00 |
| G000.3-01.6 | PHR1752-2930 | MASH1 | 17 52 52.1 | -29 30 01 | 48 | 7.80E+02 | 3.03E+00 | 2.41 | 0.00 |
| G000.4-01.9 | M 2-20 | ACKER | 17 54 25.32 | -29 36 09.4 | 52 | 2.99E+03 | 8.30E+00 | 0.95 | 0.00 |
| G000.4-02.9 | M 3-19 | ACKER | 17 58 19.47 | -30 00 41.3 | 32 | 9.31E+02 | 3.26E+00 | 2.21 | 0.00 |
| G000.5+02.8 | PHR1735-2659 | MASH1 | 17 35 55.6 | -26 59 17 | 48 | 1.80E+02 | 4.39E+00 | 4.00 | 0.03 |
| G000.5-01.6 | Al 2-Q | ACKER | 17 53 24.93 | -29 17 06.8 | 34 | 6.02E+02 | 3.53E+00 | 2.69 | 0.01 |
| G000.6-01.4 | PHR1753-2905 | MASH1 | 17 53 00.7 | -29 05 53 | 40 | 1.77E+02 | 4.04E+00 | 4.02 | 0.02 |
| G000.6-02.3 | H 2-32 | ACKER | 17 56 24.18 | -29 38 04.7 | 30 | 1.40E+02 | 1.92E+00 | 4.27 | 0.01 |
| G000.7-02.7 | M 2-21 | ACKER | 17 58 09.75 | -29 44 21.0 | 34 | 1.11E+03 | 1.71E+00 | 2.02 | 0.00 |
| G000.8-01.5 | Bl O | ACKER | 17 53 50.78 | -28 59 21.9 | 64 | 3.27E+03 | 2.07E+01 | 0.85 | 0.01 |
| G000.9+02.1 | MPA1739-2702 | MASH2 | 17 39 30.1 | -27 02 02 | 44 | 3.07E+02 | 3.46E+00 | 3.42 | 0.01 |
| G000.9-01.0 | MPA1751-2838 | MASH2 | 17 51 43.3 | -28 38 59 | 40 | 6.63E+02 | 1.67E+01 | 2.58 | 0.03 |
| G000.9-02.0 | Bl 3-13 | ACKER | 17 56 01.79 | -29 11 17.3 | 30 | 1.23E+03 | 3.10E+00 | 1.91 | 0.00 |
| G001.0+01.9 | K 1- 4 | ACKER | 17 40 28.09 | -27 00 47.1 | 32 | 1.75E+02 | 3.17E+00 | 4.03 | 0.02 |
| G001.0-01.9 | PHR1755-2904 | MASH1 | 17 55 43.1 | -29 04 05 | 56 | 1.36E+03 | 8.42E+00 | 1.80 | 0.01 |
| G001.0-02.6 | Sa 3-104 | ACKER | 17 58 25.35 | -29 20 49.9 | 68 | 2.81E+03 | 2.23E+01 | 1.01 | 0.01 |
| G001.1+02.2 | MPA1739-2648 | MASH2 | 17 39 49.7 | -26 48 45 | 30 | 1.52E+02 | 2.22E+00 | 4.18 | 0.02 |
| G001.1-01.6 | Sa 3- 92 | ACKER | 17 54 51.92 | -28 48 55.4 | 32 | 1.34E+02 | 3.94E+00 | 4.32 | 0.03 |
| G001.3-01.2 | Bl M | ACKER | 17 53 46.97 | -28 27 16.1 | 60 | 2.50E+03 | 1.42E+01 | 1.14 | 0.01 |
| G001.5-02.4 | PHR1758-2852 | MASH1 | 17 58 49.2 | -28 52 56 | 40 | 3.38E+01 | 4.79E+00 | 5.81 | 0.16 |
| G001.6-00.6 | PHR1751-2748 | MASH1 | 17 51 55.7 | -27 48 11 | 30 | 1.46E+02 | 1.32E+01 | 4.23 | 0.10 |
| G001.6-01.3 | Bl Q | ACKER | 17 54 34.61 | -28 12 42.6 | 68 | 2.73E+03 | 1.01E+02 | 1.05 | 0.04 |
| G001.6-02.6 | PHR1759-2853 | MASH1 | 17 59 56.8 | -28 53 55 | 50 | 1.36E+02 | 1.19E+01 | 4.30 | 0.09 |
| G001.7-01.6 | H 2-31 | ACKER | 17 56 02.65 | -28 14 12.2 | 60 | 1.79E+03 | 1.76E+01 | 1.50 | 0.01 |
| G001.7-02.6 | PPA1800-2846 | MASH1 | 18 00 00.6 | -28 46 27 | 50 | 3.70E+02 | 4.87E+00 | 3.22 | 0.01 |

| PNG | NAME | CAT. | R.A. (2000) H:M:S | DEC (2000) D:M:S | B arcsec | FLUX(24) mJy | $\sigma$(24) mJy | [24] mag | $\sigma$([24]) mag |
|---|---|---|---|---|---|---|---|---|---|
| G001.9+02.1 | PHR1741-2609 | MASH1 | 17 41 44.7 | -26 09 20 | 40 | 8.15E+01 | 7.30E+00 | 4.86 | 0.10 |
| G002.0+01.5 | PHR1744-2624 | MASH1 | 17 44 33.5 | -26 24 54 | 40 | 4.03E+01 | 1.36E+01 | 5.62 | 0.38 |
| G002.0+00.7 | MPA1747-2649 | MASH2 | 17 47 28.3 | -26 49 48 | 30 | 1.16E+03 | 1.14E+01 | 1.98 | 0.01 |
| G002.0-02.0 | H 1-45 | ACKER | 17 58 21.68 | -28 14 50.1 | 28 | 1.53E+03 | 5.20E+00 | 1.67 | 0.00 |
| G002.0-02.5 | PPA1800-2826 | MASH1 | 18 00 18.7 | -28 26 08 | 44 | 3.29E+02 | 5.30E+00 | 3.34 | 0.02 |
| G002.0-02.4 | MPA1800-2825 | MASH2 | 18 00 09.7 | -28 25 25 | 40 | 2.90E+01 | 3.25E+00 | 5.98 | 0.12 |
| G002.1+02.6 | PPA1740-2544 | MASH1 | 17 40 30.7 | -25 44 40 | 40 | 3.94E+01 | 1.05E+01 | 5.65 | 0.30 |
| G002.1-01.1 | MPA1755-2741 | MASH2 | 17 55 10.4 | -27 41 40 | 48 | 6.56E+02 | 1.58E+01 | 2.59 | 0.03 |
| G002.1-02.2 | M 3-20 | ACKER | 17 59 19.35 | -28 13 47.9 | 60 | 1.49E+03 | 2.34E+01 | 1.70 | 0.02 |
| G002.1-02.4 | PPA1800-2818 | MASH1 | 18 00 18.8 | -28 18 35 | 40 | 1.05E+02 | 6.59E+00 | 4.59 | 0.07 |
| G002.1-02.8 | PHR1801-2831 | MASH1 | 18 01 32.3 | -28 31 45 | 40 | 1.29E+02 | 2.99E+00 | 4.36 | 0.03 |
| G002.2+00.5 | Te 2337 | ACKER | 17 48 45.66 | -26 43 30.7 | 52 | 2.20E+03 | 2.10E+01 | 1.28 | 0.01 |
| G002.2-01.2 | PPA1755-2739 | MASH1 | 17 55 45.6 | -27 39 41 | 48 | 5.32E+02 | 1.90E+01 | 2.82 | 0.04 |
| G002.2-02.5 | KFL 2 | ACKER | 18 00 59.82 | -28 16 18.6 | 32 | 1.61E+02 | 5.48E+00 | 4.12 | 0.04 |
| G002.2-02.7 | M 2-23 | ACKER | 18 01 42.66 | -28 25 45.4 | 72 | 5.48E+03 | 1.08E+01 | 0.29 | 0.00 |
| G002.3+02.4 | PPA1741-2538 | MASH1 | 17 41 48.4 | -25 38 18 | 30 | 4.71E+01 | 2.04E+00 | 5.45 | 0.05 |
| G002.3+02.2 | Te 5 | ACKER | 17 42 30.10 | -25 45 28.7 | 34 | 2.20E+02 | 1.78E+00 | 3.78 | 0.01 |
| G002.3+01.7 | PHR1744-2603 | MASH1 | 17 44 35.4 | -26 03 36 | 40 | 1.99E+01 | 1.05E+01 | 6.39 | 0.64 |
| G002.3-01.9 | PHR1758-2756 | MASH1 | 17 58 35.1 | -27 56 56 | 40 | 7.43E+01 | 7.01E+00 | 4.96 | 0.10 |
| G002.4+01.1 | PHR1746-2611 | MASH1 | 17 46 52.0 | -26 11 50 | 40 | 9.06E+01 | 7.95E+00 | 4.74 | 0.10 |
| G002.5-01.7 | Pe 2-11 | ACKER | 17 58 31.24 | -27 37 05.4 | 30 | 9.24E+01 | 2.60E+00 | 4.72 | 0.03 |
| G002.6+02.3 | PHR1742-2525 | MASH1 | 17 42 52.1 | -25 25 43 | 30 | 7.44E+01 | 2.14E+00 | 4.96 | 0.03 |
| G002.6+02.1 | Te 1580 | ACKER | 17 43 39.50 | -25 36 45.0 | 40 | 1.31E+03 | 1.09E+01 | 1.85 | 0.01 |
| G002.7+01.7 | PPA1745-2542 | MASH1 | 17 45 18.9 | -25 42 05 | 30 | 6.96E+01 | 2.04E+00 | 5.03 | 0.03 |
| G002.8+01.8 | Te 1567 | ACKER | 17 45 28.36 | -25 38 13.4 | 40 | 1.35E+02 | 6.17E+00 | 4.31 | 0.05 |
| G002.8-02.2 | Pe 2-12 | ACKER | 18 01 09.78 | -27 38 25.8 | 34 | 9.26E+02 | 3.97E+00 | 2.22 | 0.00 |
| G003.0-02.6 | KFL 4 | ACKER | 18 02 51.54 | -27 41 02.4 | 26 | 4.97E+01 | 1.52E+00 | 5.39 | 0.03 |
| G003.1-01.6 | PHR1759-2706 | MASH1 | 17 59 26.1 | -27 06 34 | 40 | 7.21E+02 | 9.08E+00 | 2.49 | 0.01 |
| G003.3-01.6 | PHR1759-2651 | MASH1 | 17 59 55.3 | -26 51 49 | 50 | 5.00E+02 | 1.92E+01 | 2.89 | 0.04 |
| G003.4-01.8 | PHR1800-2653 | MASH1 | 18 00 42.3 | -26 53 37 | 40 | 8.96E+01 | 1.79E+01 | 4.75 | 0.22 |

| PNG | NAME | CAT. | R.A. (2000) H:M:S | DEC (2000) D:M:S | B arcsec | FLUX(24) mJy | $\sigma(24)$ mJy | [24] mag | $\sigma([24])$ mag |
|---|---|---|---|---|---|---|---|---|---|
| G003.5+01.3 | MPA1748-2511 | MASH2 | 17 48 41.7 | -25 11 34 | 46 | 2.43E+03 | 1.75E+01 | 1.17 | 0.01 |
| G003.5-02.4 | IC 4673 | ACKER | 18 03 17.98 | -27 06 16.4 | 86 | 6.87E+03 | 4.75E+01 | 0.04 | 0.01 |
| G003.6-01.3 | PHR1759-2630 | MASH1 | 17 59 12.1 | -26 30 24 | 44 | 1.33E+03 | 2.04E+01 | 1.83 | 0.02 |
| G003.6-02.3 | M 2-26 | ACKER | 18 03 11.99 | -26 58 30.9 | 36 | 3.50E+02 | 5.76E+00 | 3.28 | 0.02 |
| G003.9+01.6 | Te 2111 | ACKER | 17 48 28.34 | -24 41 28.9 | 72 | 1.21E+03 | 6.50E+01 | 1.93 | 0.06 |
| G003.9-02.3 | M 1-35 | ACKER | 18 03 39.33 | -26 43 24.8 | 72 | 4.84E+03 | 1.30E+01 | 0.42 | 0.00 |
| G004.0+02.6 | PHR1744-2406 | MASH1 | 17 44 46.3 | -24 06 59 | 48 | 5.01E+01 | 5.48E+00 | 5.39 | 0.12 |
| G004.0-02.6 | PHR1804-2645 | MASH1 | 18 04 59.5 | -26 45 17 | 50 | 1.61E+02 | 1.42E+01 | 4.12 | 0.10 |
| G004.0-02.7 | PPA1805-2649 | MASH1 | 18 05 26.3 | -26 49 03 | 50 | 7.26E+01 | 7.56E+00 | 4.98 | 0.11 |
| G004.2-02.5 | PHR1805-2631 | MASH1 | 18 05 20.1 | -26 31 45 | 46 | 5.89E+02 | 7.70E+00 | 2.71 | 0.01 |
| G004.3+01.8a | PHR1748-2417 | MASH1 | 17 48 33.0 | -24 17 36 | 52 | 3.51E+02 | 1.57E+01 | 3.27 | 0.05 |
| G004.3+01.8 | H 2-24 | ACKER | 17 48 36.36 | -24 16 35.5 | 72 | 3.12E+03 | 2.35E+01 | 0.90 | 0.01 |
| G004.3-01.4 | PPA1801-2553 | MASH1 | 18 01 18.9 | -25 53 21 | 50 | 1.74E+03 | 1.40E+01 | 1.53 | 0.01 |
| G004.3-02.6 | H 1-53 | ACKER | 18 05 57.09 | -26 29 41.8 | 32 | 9.72E+02 | 2.99E+00 | 2.17 | 0.00 |
| G004.5+02.0 | MPA1748-2402 | MASH2 | 17 48 21.4 | -24 02 14 | 50 | 3.28E+02 | 5.69E+00 | 3.35 | 0.02 |
| G004.6+01.8 | BMP1749-2356 | MASH2 | 17 49 14.8 | -23 56 42 | 40 | 1.02E+02 | 1.73E+01 | 4.61 | 0.19 |
| G004.8-01.1 | PHR1801-2522 | MASH1 | 18 01 16.9 | -25 22 38 | 80 | 6.78E+03 | 7.99E+01 | 0.06 | 0.01 |
| G005.0+02.2 | PHR1748-2326 | MASH1 | 17 48 46.0 | -23 26 27 | 36 | 9.10E+01 | 8.75E+00 | 4.74 | 0.10 |
| G005.2-01.6 | PPA1803-2516 | MASH1 | 18 03 52.4 | -25 16 59 | 40 | 1.41E+02 | 1.05E+01 | 4.26 | 0.08 |
| G005.2-02.4 | PHR1807-2535 | MASH1 | 18 07 03.3 | -25 35 43 | 56 | 4.71E+02 | 2.29E+01 | 2.95 | 0.05 |
| G005.5+02.7 | H 1-34 | ACKER | 17 48 07.78 | -22 46 46.5 | 74 | 5.21E+03 | 3.10E+01 | 0.34 | 0.01 |
| G005.5-02.5 | M 3-24 | ACKER | 18 07 53.51 | -25 24 02.2 | 36 | 1.05E+03 | 9.13E+00 | 2.08 | 0.01 |
| G005.8+02.2a | MPA1750-2248a | MASH2 | 17 50 20.7 | -22 48 24 | 48 | 3.99E+02 | 1.21E+01 | 3.13 | 0.03 |
| G005.9-02.6 | MaC 1-10 | ACKER | 18 09 12.54 | -25 04 35.5 | 68 | 9.90E+03 | 4.52E+01 | -0.35 | 0.00 |
| G006.0+02.8 | Th 4- 3 | ACKER | 17 48 37.11 | -22 16 49.3 | 38 | 1.71E+03 | 3.56E+00 | 1.55 | 0.00 |
| G006.1+01.5 | PHR1753-2254 | MASH1 | 17 53 45.3 | -22 54 01 | 80 | 3.15E+03 | 3.10E+01 | 0.89 | 0.01 |
| G006.1+00.8 | PPA1756-2311 | MASH1 | 17 56 33.2 | -23 11 47 | 48 | 6.61E+02 | 1.43E+01 | 2.59 | 0.02 |
| G006.1-02.1 | PPA1807-2439 | MASH1 | 18 07 40.9 | -24 39 18 | 40 | 2.02E+02 | 1.94E+01 | 3.87 | 0.10 |
| G006.2+01.0 | HaTr 8 | ACKER | 17 55 55.91 | -22 59 01.4 | 36 | 1.51E+02 | 5.21E+00 | 4.19 | 0.04 |
| G006.3+02.2 | MPA1751-2223 | MASH2 | 17 51 40.0 | -22 23 18 | 46 | 2.65E+02 | 9.63E+00 | 3.58 | 0.04 |

| PNG | NAME | CAT. | R.A. (2000) H:M:S | DEC (2000) D:M:S | B arcsec | FLUX(24) mJy | $\sigma(24)$ mJy | [24] mag | $\sigma([24])$ mag |
|---|---|---|---|---|---|---|---|---|---|
| G006.4+02.0 | M 1-31 | ACKER | 17 52 41.45 | -22 21 56.6 | 64 | 6.63E+03 | 2.01E+01 | 0.08 | 0.00 |
| G006.7-02.2 | M 1-41 | ACKER | 18 09 29.91 | -24 12 28.2 | 80 | 1.25E+04 | 1.13E+02 | -0.61 | 0.01 |
| G006.8+02.0 | Pe 2-10 | ACKER | 17 53 36.73 | -21 58 40.5 | 36 | 1.85E+02 | 3.19E+00 | 3.97 | 0.02 |
| G006.8+02.3 | Th 4- 7 | ACKER | 17 52 22.55 | -21 51 12.9 | 36 | 6.16E+02 | 2.61E+00 | 2.66 | 0.00 |
| G006.9+01.5 | MPA1755-2212 | MASH2 | 17 55 36.7 | -22 12 47 | 50 | 4.46E+02 | 4.31E+00 | 3.01 | 0.01 |
| G007.2+01.8 | Hb 6 | ACKER | 17 55 07.22 | -21 44 37.9 | 68 | 1.17E+04 | 3.29E+01 | -0.54 | 0.00 |
| G007.3+01.7 | PHR1755-2142 | MASH1 | 17 55 34.7 | -21 42 38 | 50 | 4.74E+02 | 9.56E+00 | 2.95 | 0.02 |
| G007.4+01.7 | PHR1755-2140 | MASH1 | 17 55 42.6 | -21 40 18 | 44 | 3.91E+02 | 4.64E+00 | 3.16 | 0.01 |
| G007.6+02.0 | PHR1755-2118 | MASH1 | 17 55 03.6 | -21 18 38 | 44 | 1.26E+02 | 7.50E+00 | 4.38 | 0.06 |
| G007.7-01.6 | PHR1809-2303 | MASH1 | 18 09 12.7 | -23 03 58 | 40 | 5.22E+02 | 7.58E+00 | 2.84 | 0.02 |
| G008.3-01.1 | M 1-40 | ACKER | 18 08 25.36 | -22 16 52.8 | 70 | 1.33E+04 | 8.08E+01 | -0.67 | 0.01 |
| G010.1+00.7 | NGC 6537 | ACKER | 18 05 13.39 | -19 50 13.6 | 70 | 2.68E+04 | 5.31E+01 | -1.43 | 0.00 |
| G011.7-00.6 | NGC 6567 | ACKER | 18 13 45.06 | -19 04 19.3 | 40 | 4.80E+03 | 2.10E+01 | 0.43 | 0.00 |
| G019.1+00.8 | MPA1822-1153 | MASH2 | 18 22 53.7 | -11 53 09 | 50 | 1.55E+03 | 2.14E+01 | 1.66 | 0.01 |
| G019.6+00.7 | MPA1824-1126 | MASH2 | 18 24 04.0 | -11 26 15 | 50 | 2.92E+02 | 2.07E+01 | 3.47 | 0.08 |
| G019.9+00.9 | M 3-53 | ACKER | 18 24 07.86 | -11 06 44.7 | 40 | 1.84E+03 | 9.40E+00 | 1.47 | 0.01 |
| G021.7-00.6 | M 3-55 | ACKER | 18 33 14.67 | -10 15 07.1 | 30 | 1.46E+02 | 2.20E+01 | 4.22 | 0.16 |
| G021.8-00.4 | M 3-28 | ACKER | 18 32 41.19 | -10 05 48.5 | 32 | 1.26E+03 | 1.99E+01 | 1.88 | 0.02 |
| G023.4+00.7 | PHR1831-0805 | MASH1 | 18 31 19.6 | -08 05 43 | 44 | 2.41E+02 | 2.45E+01 | 3.68 | 0.11 |
| G024.4+00.9 | MPA1832-0706 | MASH2 | 18 32 22.9 | -07 06 57 | 50 | 1.64E+03 | 2.53E+01 | 1.60 | 0.02 |
| G026.8-00.1 | MPA1840-0529 | MASH2 | 18 40 49.3 | -05 29 44 | 40 | 7.53E+02 | 8.01E+01 | 2.44 | 0.12 |
| G027.6-00.8 | PHR1844-0503 | MASH1 | 18 44 45.7 | -05 03 54 | 44 | 6.87E+02 | 3.51E+01 | 2.54 | 0.06 |
| G027.7+00.7 | M 2-45 | ACKER | 18 39 21.80 | -04 19 50.4 | 68 | 8.17E+03 | 6.88E+01 | -0.14 | 0.01 |
| G027.8+00.5 | MPA1840-0415 | MASH2 | 18 40 25.3 | -04 15 34 | 24 | 6.95E+01 | 5.97E+00 | 5.03 | 0.09 |
| G027.8-00.7 | MPA1844-0454 | MASH2 | 18 44 49.6 | -04 54 00 | 40 | 1.71E+02 | 2.48E+01 | 4.05 | 0.16 |
| G028.9+00.2 | PHR1843-0325 | MASH1 | 18 43 15.3 | -03 25 27 | 44 | 9.36E+02 | 1.20E+02 | 2.21 | 0.14 |
| G029.0+00.4 | A 48 | ACKER | 18 42 49.51 | -03 12 59.4 | 50 | 2.03E+03 | 1.22E+02 | 1.37 | 0.07 |
| G029.2-00.0 | TDC 1 | ACKER | 18 44 53.45 | -03 20 33.5 | 38 | 8.07E+03 | 6.63E+02 | -0.13 | 0.09 |
| G031.3-00.5 | HaTr 10 | ACKER | 18 50 24.72 | -01 40 08.8 | 30 | 1.43E+02 | 1.07E+01 | 4.25 | 0.08 |
| G031.9-00.3 | WeSb 4 | ACKER | 18 50 40.22 | -01 03 13.7 | 42 | 1.38E+03 | 8.89E+01 | 1.78 | 0.07 |

| PNG | NAME | CAT. | R.A. (2000) H:M:S | DEC (2000) D:M:S | B arcsec | FLUX(24) mJy | $\sigma(24)$ mJy | [24] mag | $\sigma$([24]) mag |
|---|---|---|---|---|---|---|---|---|---|
| G032.3-00.5 | MPA1852-0044 | MASH2 | 18 52 24.1 | -00 44 46 | 44 | 2.24E+02 | 2.70E+01 | 3.76 | 0.13 |
| G032.5-00.3 | MPA1851-0028 | MASH2 | 18 51 47.5 | -00 28 29 | 40 | 1.25E+02 | 3.90E+01 | 4.39 | 0.35 |
| G040.3-00.4 | A 53 | ACKER | 19 06 45.84 | +06 23 55.8 | 50 | 1.18E+03 | 2.55E+01 | 1.95 | 0.02 |
| G041.2-00.6 | HaTr 14 | ACKER | 19 09 13.48 | +07 05 43.1 | 36 | 8.16E+01 | 1.52E+01 | 4.86 | 0.20 |
| G055.5-00.5 | M 1-71 | ACKER | 19 36 26.56 | +19 42 30.0 | 74 | 1.35E+04 | 7.52E+01 | -0.69 | 0.01 |
| G056.4-00.9 | K 3-42 | ACKER | 19 39 35.77 | +20 19 02.0 | 36 | 1.80E+03 | 2.44E+00 | 1.50 | 0.00 |
| G060.5-00.3 | K 3-45 | ACKER | 19 46 15.66 | +24 11 07.1 | 32 | 8.48E+01 | 5.59E+00 | 4.81 | 0.07 |
| G062.4-00.2 | M 2-48 | ACKER | 19 50 28.14 | +25 54 21.6 | 36 | 4.64E+02 | 3.64E+00 | 2.97 | 0.01 |
| G065.9+00.5 | NGC 6842 | ACKER | 19 55 02.34 | +29 17 21.3 | 74 | 1.24E+03 | 3.16E+01 | 1.90 | 0.03 |
| G300.2+00.6 | He 2- 83 | ACKER | 12 28 45.69 | -62 05 35.4 | 72 | 8.32E+03 | 5.29E+01 | -0.16 | 0.01 |
| G300.4-00.9 | He 2- 84 | ACKER | 12 28 46.79 | -63 44 35.4 | 36 | 2.93E+02 | 2.59E+00 | 3.47 | 0.01 |
| G301.1-00.4 | MPA1235-6318 | MASH2 | 12 35 21.4 | -63 18 01 | 50 | 1.98E+02 | 9.88E+00 | 3.89 | 0.05 |
| G302.3-00.5 | PHR1246-6324 | MASH1 | 12 46 26.5 | -63 24 28 | 54 | 9.56E+02 | 6.64E+00 | 2.18 | 0.01 |
| G302.6-00.9 | Wray 16-121 | ACKER | 12 48 31.09 | -63 49 56.5 | 50 | 6.88E+02 | 4.37E+01 | 2.54 | 0.07 |
| G305.6-00.9 | MPA1315-6338 | MASH2 | 13 15 30.4 | -63 38 43 | 50 | 5.73E+02 | 1.80E+01 | 2.74 | 0.03 |
| G306.4-00.6 | Th 2-A | ACKER | 13 22 34.84 | -63 20 55.7 | 52 | 2.61E+03 | 1.75E+01 | 1.10 | 0.01 |
| G308.1-00.5 | MPA1337-6258 | MASH2 | 13 37 54.9 | -62 58 54 | 50 | 9.29E+01 | 4.39E+00 | 4.72 | 0.05 |
| G309.0+00.8 | He 2- 96 | ACKER | 13 42 35.67 | -61 22 31.0 | 68 | 8.08E+03 | 4.22E+01 | -0.13 | 0.01 |
| G315.0-00.3 | He 2-111 | ACKER | 14 33 18.30 | -60 49 44.6 | 70 | 3.33E+03 | 5.62E+01 | 0.83 | 0.02 |
| G318.9+00.7 | PHR1457-5812 | MASH1 | 14 57 35.7 | -58 12 09 | 60 | 1.80E+03 | 1.92E+01 | 1.50 | 0.01 |
| G322.2-00.4 | BMP1522-5729 | MASH2 | 15 22 58.9 | -57 29 59 | 50 | 1.67E+03 | 2.01E+01 | 1.58 | 0.01 |
| G322.2-00.7 | BMP1524-5746 | MASH2 | 15 24 24.0 | -57 46 21 | 80 | 4.25E+03 | 4.03E+01 | 0.56 | 0.01 |
| G322.4-00.1 | MPA1523-5710 | MASH2 | 15 23 22.5 | -57 10 48 | 50 | 7.83E+02 | 4.13E+01 | 2.40 | 0.06 |
| G322.4-00.1 | Pe 2- 8 | ACKER | 15 23 42.21 | -57 09 17.5 | 72 | 2.32E+04 | 1.04E+02 | -1.28 | 0.00 |
| G329.5-00.8 | MPA1605-5319 | MASH2 | 16 05 37.4 | -53 19 54 | 50 | 6.05E+02 | 2.56E+01 | 2.68 | 0.05 |
| G331.4+00.5 | He 2-145 | ACKER | 16 08 59.41 | -51 02 02.1 | 42 | 6.31E+02 | 3.38E+02 | 2.64 | 0.65 |
| G333.9+00.6 | PHR1619-4914 | MASH1 | 16 19 40.1 | -49 14 00 | 240 | 2.69E+04 | 5.39E+01 | -1.44 | 0.00 |
| G337.3+00.6 | PHR1633-4650 | MASH1 | 16 33 58.0 | -46 50 07 | 44 | 1.48E+03 | 4.66E+01 | 1.71 | 0.03 |
| G337.6+00.7 | PHR1634-4628 | MASH1 | 16 34 51.2 | -46 28 28 | 32 | 1.13E+02 | 1.04E+01 | 4.50 | 0.10 |

| PNG | NAME | CAT. | R.A. (2000) H:M:S | DEC (2000) D:M:S | B arcsec | FLUX(24) mJy | σ(24) mJy | [24] mag | σ([24]) mag |
|---|---|---|---|---|---|---|---|---|---|
| G339.1+00.9 | PHR1639-4516 | MASH1 | 16 39 22.3 | -45 16 35 | 48 | 4.59E+01 | 2.40E+01 | 5.48 | 0.63 |
| G342.7+00.7 | H 1- 3 | ACKER | 16 53 31.55 | -42 39 18.1 | 40 | 5.49E+02 | 1.41E+01 | 2.79 | 0.03 |
| G343.9+00.8 | H 1- 5 | ACKER | 16 57 23.77 | -41 37 55.9 | 72 | 1.76E+04 | 5.37E+01 | -0.98 | 0.00 |
| G345.4+00.1 | IC 4637 | ACKER | 17 05 09.01 | -40 52 57.1 | 58 | 1.36E+04 | 6.59E+01 | -0.70 | 0.01 |
| G346.8-00.7 | MPA1713-4015 | MASH2 | 17 13 10.8 | -40 15 56 | 50 | 1.49E+03 | 1.22E+01 | 1.71 | 0.01 |
| G347.2-00.8 | PHR1714-4006 | MASH1 | 17 14 49.3 | -40 06 09 | 40 | 1.54E+02 | 2.87E+01 | 4.17 | 0.20 |
| G352.4-02.7 | PPA1737-3650 | MASH1 | 17 37 37.6 | -36 50 18 | 36 | 2.99E+02 | 3.15E+00 | 3.45 | 0.01 |
| G352.8-00.2 | H 1-13 | ACKER | 17 28 27.27 | -35 07 42.8 | 80 | 2.13E+04 | 2.00E+02 | -1.19 | 0.01 |
| G352.8-00.5 | MPA1729-3513 | MASH2 | 17 29 37.5 | -35 13 46 | 72 | 1.20E+04 | 2.49E+02 | -0.56 | 0.02 |
| G353.3-02.2 | PPA1738-3546 | MASH1 | 17 38 16.2 | -35 46 29 | 52 | 6.09E+02 | 1.09E+01 | 2.67 | 0.02 |
| G353.3-02.9 | PHR1740-3607 | MASH1 | 17 40 50.3 | -36 07 45 | 40 | 8.71E+01 | 4.25E+00 | 4.79 | 0.05 |
| G353.5-02.6 | PPA1740-3551 | MASH1 | 17 40 21.4 | -35 51 31 | 46 | 3.42E+02 | 1.78E+01 | 3.30 | 0.06 |
| G353.6+01.7 | PPA1722-3317 | MASH1 | 17 22 35.4 | -33 17 15 | 54 | 3.07E+03 | 2.03E+01 | 0.92 | 0.01 |
| G354.4+02.2 | PPA1723-3223 | MASH1 | 17 23 04.2 | -32 23 10 | 50 | 8.05E+02 | 6.79E+00 | 2.37 | 0.01 |
| G354.5+02.4 | PHR1722-3210 | MASH1 | 17 22 11.7 | -32 10 45 | 50 | 6.37E+02 | 3.57E+00 | 2.63 | 0.01 |
| G354.5-02.0a | PPA1740-3437 | MASH1 | 17 40 30.5 | -34 37 17 | 48 | 3.79E+02 | 1.22E+01 | 3.19 | 0.03 |
| G354.6-01.4 | PPA1737-3414 | MASH1 | 17 37 53.9 | -34 14 27 | 48 | 5.00E+02 | 1.55E+01 | 2.89 | 0.03 |
| G354.7+02.8 | PPA1721-3149 | MASH1 | 17 21 23.7 | -31 49 52 | 48 | 4.25E+02 | 1.35E+01 | 3.07 | 0.03 |
| G354.8+01.8 | PPA1725-3216 | MASH1 | 17 25 15.9 | -32 16 11 | 48 | 1.08E+03 | 1.23E+01 | 2.06 | 0.01 |
| G354.9-02.8 | MPA1744-3444 | MASH2 | 17 44 43.5 | -34 44 26 | 40 | 2.42E+01 | 1.89E+00 | 6.18 | 0.08 |
| G355.0+02.6 | PPA1722-3139 | MASH1 | 17 22 40.8 | -31 39 55 | 50 | 4.23E+03 | 2.60E+01 | 0.57 | 0.01 |
| G355.1+02.3 | Th 3-11 | ACKER | 17 24 26.51 | -31 43 17.9 | 40 | 2.85E+03 | 1.55E+01 | 1.00 | 0.01 |
| G355.2-02.0 | PPA1741-3405 | MASH1 | 17 41 59.1 | -34 05 34 | 46 | 3.83E+02 | 5.79E+00 | 3.18 | 0.02 |
| G355.2-02.5 | H 1-29 | ACKER | 17 44 14.20 | -34 17 28.1 | 36 | 1.17E+03 | 2.63E+00 | 1.96 | 0.00 |
| G355.4-02.4 | M 3-14 | ACKER | 17 44 20.50 | -34 06 40.7 | 40 | 2.64E+03 | 4.49E+00 | 1.08 | 0.00 |
| G355.6-01.4 | PHR1740-3324 | MASH1 | 17 40 54.9 | -33 24 18 | 50 | 3.30E+02 | 3.23E+01 | 3.34 | 0.11 |
| G355.8+01.7 | MPA1728-3132 | MASH2 | 17 28 31.1 | -31 32 09 | 50 | 8.33E+02 | 2.50E+01 | 2.33 | 0.03 |
| G355.9+02.7 | Th 3-10 | ACKER | 17 24 40.17 | -30 51 52.9 | 40 | 2.47E+03 | 4.92E+02 | 1.16 | 0.22 |
| G356.0+02.8 | PPA1724-3043 | MASH1 | 17 24 58.3 | -30 43 04 | 40 | 9.48E+02 | 7.42E+00 | 2.19 | 0.01 |

| PNG | NAME | CAT. | R.A. (2000) H:M:S | DEC (2000) D:M:S | B arcsec | FLUX(24) mJy | σ(24) mJy | [24] mag | σ([24]) mag |
|---|---|---|---|---|---|---|---|---|---|
| G356.0-01.4 | PPA1741-3302 | MASH1 | 17 41 33.4 | -33 02 15 | 48 | 6.83E+02 | 1.26E+01 | 2.55 | 0.02 |
| G356.0-01.8 | PPA1743-3315 | MASH1 | 17 43 07.2 | -33 15 54 | 48 | 7.19E+02 | 2.44E+01 | 2.49 | 0.04 |
| G356.1+02.7 | Th 3-13 | ACKER | 17 25 18.89 | -30 40 44.1 | 70 | 5.59E+03 | 2.56E+01 | 0.27 | 0.00 |
| G356.1-02.7 | PPA1747-3341 | MASH1 | 17 47 04.8 | -33 41 03 | 50 | 6.92E+02 | 3.26E+01 | 2.53 | 0.05 |
| G356.2+02.7 | MPA1725-3033 | MASH2 | 17 25 33.4 | -30 33 57 | 40 | 6.23E+01 | 4.40E+00 | 5.15 | 0.08 |
| G356.2+02.5 | PPA1726-3045 | MASH1 | 17 26 23.6 | -30 45 39 | 40 | 5.03E+02 | 6.35E+00 | 2.88 | 0.01 |
| G356.3-02.6 | MPA1747-3326 | MASH2 | 17 47 27.5 | -33 26 38 | 46 | 5.32E+01 | 1.19E+01 | 5.32 | 0.25 |
| G356.5+02.2 | PHR1728-3038 | MASH1 | 17 28 07.7 | -30 38 18 | 46 | 4.21E+02 | 8.35E+00 | 3.07 | 0.02 |
| G356.5+01.5 | Th 3-55 | ACKER | 17 30 58.58 | -31 01 07.6 | 40 | 2.13E+03 | 1.41E+01 | 1.32 | 0.01 |
| G356.5-02.3 | M 1-27 | ACKER | 17 46 45.51 | -33 08 35.1 | 74 | 1.32E+04 | 7.26E+00 | -0.67 | 0.00 |
| G356.6+02.3 | PHR1728-3032 | MASH1 | 17 28 14.2 | -30 32 14 | 46 | 2.49E+02 | 6.99E+00 | 3.65 | 0.03 |
| G356.6-01.9 | PHR1745-3246 | MASH1 | 17 45 09.8 | -32 46 17 | 110 | 6.58E+03 | 1.30E+02 | 0.09 | 0.02 |
| G356.9+00.9 | PPA1734-3102 | MASH1 | 17 34 34.3 | -31 02 08 | 70 | 4.51E+03 | 1.73E+01 | 0.50 | 0.00 |
| G357.0+02.4 | M 4- 4 | ACKER | 17 28 50.14 | -30 07 52.8 | 40 | 1.06E+03 | 7.25E+00 | 2.08 | 0.01 |
| G357.1+01.9 | Th 3-24 | ACKER | 17 30 51.43 | -30 17 14.1 | 34 | 1.31E+02 | 9.10E+00 | 4.34 | 0.08 |
| G357.2+01.4 | Al 2-H | ACKER | 17 33 17.03 | -30 26 30.6 | 40 | 6.25E+02 | 1.10E+01 | 2.65 | 0.02 |
| G357.2+02.0 | H 2-13 | ACKER | 17 31 07.27 | -30 10 36.9 | 36 | 1.19E+03 | 1.12E+01 | 1.95 | 0.01 |
| G357.3+01.3 | PHR1733-3029 | MASH1 | 17 33 58.2 | -30 29 46 | 52 | 1.48E+02 | 3.38E+01 | 4.21 | 0.25 |
| G357.3-02.0 | PPA1747-3215 | MASH1 | 17 47 28.5 | -32 15 46 | 50 | 2.16E+03 | 1.71E+01 | 1.30 | 0.01 |
| G357.5-02.4 | PPA1749-3216 | MASH1 | 17 49 37.9 | -32 16 28 | 48 | 6.44E+02 | 4.29E+00 | 2.61 | 0.01 |
| G357.6+01.0 | TrBr 4 | ACKER | 17 35 43.65 | -30 21 29.0 | 46 | 5.54E+02 | 2.21E+01 | 2.78 | 0.04 |
| G357.6+01.7 | H 1-23 | ACKER | 17 32 46.94 | -30 00 14.9 | 40 | 3.53E+03 | 1.43E+01 | 0.77 | 0.00 |
| G357.6+02.6 | H 1-18 | ACKER | 17 29 43.27 | -29 32 49.0 | 40 | 3.72E+03 | 1.31E+01 | 0.71 | 0.00 |
| G357.7+01.4 | PPA1734-3004 | MASH1 | 17 34 46.6 | -30 04 21 | 40 | 2.02E+02 | 3.08E+00 | 3.87 | 0.02 |
| G357.8+01.6 | PPA1734-2954 | MASH1 | 17 34 01.7 | -29 54 35 | 40 | 8.10E+02 | 2.39E+01 | 2.36 | 0.03 |
| G358.2-01.1 | Bl D | ACKER | 17 46 02.92 | -31 03 32.0 | 40 | 3.61E+02 | 1.79E+01 | 3.24 | 0.05 |
| G358.3+01.2 | Bl B | ACKER | 17 37 00.61 | -29 40 17.3 | 38 | 2.44E+03 | 1.10E+01 | 1.17 | 0.00 |
| G358.3-02.5 | Al 2-O | ACKER | 17 51 44.74 | -31 36 00.2 | 38 | 1.21E+03 | 5.77E+00 | 1.93 | 0.01 |
| G358.5+02.6 | HDW 8 | ACKER | 17 31 46.93 | -28 41 57.0 | 60 | 2.21E+03 | 1.29E+01 | 1.28 | 0.01 |

TABLE 1 (CONT.)

| PNG | NAME | CAT. | R.A. (2000) H:M:S | DEC (2000) D:M:S | B arcsec | FLUX(24) mJy | $\sigma(24)$ mJy | [24] mag | $\sigma([24])$ mag |
|---|---|---|---|---|---|---|---|---|---|
| G358.6+01.8 | M 4- 6 | ACKER | 17 35 14.03 | -29 03 11.0 | 52 | 3.55E+03 | 1.50E+01 | 0.76 | 0.00 |
| G358.7-02.5 | PHR1752-3116 | MASH1 | 17 52 36.5 | -31 16 27 | 30 | 9.61E+01 | 3.67E+00 | 4.68 | 0.04 |
| G358.7-02.7 | Al 2-R | ACKER | 17 53 37.38 | -31 25 33.0 | 40 | 6.31E+02 | 8.99E+00 | 2.64 | 0.02 |
| G359.0+02.8 | Al 2-G | ACKER | 17 32 22.56 | -28 14 30.4 | 38 | 1.06E+03 | 6.99E+00 | 2.07 | 0.01 |
| G359.1-01.7 | M 1-29 | ACKER | 17 50 18.01 | -30 34 55.5 | 48 | 6.15E+03 | 4.37E+01 | 0.16 | 0.01 |
| G359.1-02.3 | M 3-16 | ACKER | 17 52 46.62 | -30 49 44.6 | 36 | 5.61E+02 | 5.58E+00 | 2.76 | 0.01 |
| G359.1-02.9 | M 3-46 | ACKER | 17 55 06.24 | -31 12 22.5 | 40 | 4.55E+02 | 4.82E+00 | 2.99 | 0.01 |
| G359.2+01.2 | 19W32 | ACKER | 17 39 02.94 | -28 56 37.4 | 72 | 8.81E+03 | 5.54E+01 | -0.23 | 0.01 |
| G359.2-02.4 | PHR1753-3051 | MASH1 | 17 53 39.8 | -30 51 25 | 40 | 1.54E+02 | 3.90E+00 | 4.17 | 0.03 |
| G359.3+01.4 | Th 3-35 | ACKER | 17 38 42.18 | -28 42 45.9 | 40 | 3.68E+03 | 1.10E+01 | 0.72 | 0.00 |
| G359.3-01.8 | M 3-44 | ACKER | 17 51 18.93 | -30 23 53.6 | 68 | 9.28E+03 | 1.77E+01 | -0.28 | 0.00 |
| G359.4+02.3 | Th 3-32 | ACKER | 17 35 15.23 | -28 07 06.9 | 40 | 3.16E+03 | 4.42E+00 | 0.89 | 0.00 |
| G359.4+02.3a | PPA1735-2809 | MASH1 | 17 35 12.0 | -28 09 31 | 40 | 1.73E+02 | 4.55E+00 | 4.04 | 0.03 |
| G359.5+02.6 | Al 2-K | ACKER | 17 34 13.94 | -27 56 00.3 | 32 | 1.38E+03 | 4.63E+00 | 1.79 | 0.00 |
| G359.6+02.2 | Al 2-I | ACKER | 17 36 14.29 | -28 00 45.6 | 44 | 8.97E+02 | 1.00E+01 | 2.25 | 0.01 |
| G359.7-01.8 | M 3-45 | ACKER | 17 52 05.46 | -30 05 17.6 | 36 | 1.02E+03 | 3.14E+00 | 2.11 | 0.00 |
| G359.7-02.6 | H 1-40 | ACKER | 17 55 36.12 | -30 33 32.5 | 66 | 9.48E+03 | 3.56E+01 | -0.31 | 0.00 |

TABLE 2

Normalised Profile Widths for Eight Galactic Planetary Nebulae

| SOURCE | $\theta_{PNN}[3.6]$ | $\theta_{PNN}[4.5]$ | $\theta_{PNN}[5.8]$ | $\theta_{PNN}[8.0]$ | $\theta_{PNN}[24]$ | $\theta_{PN}[8.0]$ arcsec |
|---|---|---|---|---|---|---|
| NGC4673 | … | 0.94 | 1.09 | 1.00 | 1.13 | 11.6 |
| Th 2-A | 0.91 | 1.00 | 1.05 | 1.00 | 1.19 | 17.4 |
| He 2-111 | 1.02 | 1.11 | 1.45 | 1.00 | 1.58 | 10.8 |
| PHR 1246-6324 | 0.57 | 0.73 | 0.78 | 1.00 | $\leq 0.56$ | 7.56 |
| PHR 1457-5812 | 0.84 | 0.86 | 1.02 | 1.00 | 0.78 | 16.6 |
| PHR 1619-4914 | 0.92 | 0.96 | 1.00 | 1.00 | 1.05 | 32.5 |
| PPA 1755-2739 | … | 0.93 | 1.02 | 1.00 | … | 6.53 |
| PHR 1802-2522 | … | 0.93 | 1.16 | 1.00 | $\leq 0.35$ | 7.08 |
| MPA 1832-0706 | … | 1.03 | 1.09 | 1.00 | $\leq 0.44$ | 7.96 |

**Figure Captions**

**Figure 1**

Comparison of the present 24 μm fluxes for Galactic PNe with previous measurements by Zhang & Kwok (2009) and Mizuno et al. (2010). The dashed lines correspond to least-squares fits, whilst the solid diagonal lines represent one-to-one relations. It will be noted that the Mizuno et al. (2010) and Zhang & Kwok (2009) results are systematically different from those presented here, and imply smaller mean values for the 24 μm emission.

**Figure 2**

Images of 36 PNe having extended emission at 24 μm, where we have combined results at 4.5 μm (blue), 8.0 μm (green) and 24 μm (red). No shorter wave IRAC results were available for the sources PHR 1743-2431 & 1722-3210, and MPA 1748-2402 & 1739-2702, and their images correspond to 24 μm emission alone. It will be noted that the 24 μm envelopes extend, in most cases, to significant distances outside of the central cores, and usually have a circular appearance. This suggests that the longer wave fluxes arise in AGB mass-loss regimes. The white horizontal bars indicate a scale of 10 arcsec, whilst Galactic north is to the top, and east is to the left.

**Figure 3**

Profiles through six representative PNe having extended 24 μm emission, and bright IRAC cores. The dashed curves correspond to the 24 μm PSF, whilst 24 μm results are indicated using filled circles; 8.0 μm results with open diamonds; the 5.8 μm trends are indicated by open squares; 4.5 μm results by filled triangles; and 3.6 μm results by filled diamonds. It will be noticed that profile widths tend to increase with increasing MIR wavelength. All of the slices pass along an E-W direction through the centres of the sources, whilst the width of the MIPSGAL slices is 2.5 arcsec, and of the comparative IRAC profiles is 2.4 arcsec.

**Figure 4**

The logarithmic variation of 8.0 and 24 μm emission towards the westerly sectors of three PNe, where details of orientation and slice widths are as stated in Fig. 3. The solid curve corresponds to the 24 μm PSF, differing portions for which are labelled below, whilst least-squares fits in the regime 10 arcsec < r < 35 arcsec are indicated using grey dashed lines.

**Figure 5**

The location of Acker et al. (1992) (grey diamonds), MASH I (open circles), and MASH II (filled triangles) PNe in the [8.0]-[24]/[3.6]-[5.8] colour plane. Note that all of the PNe have large [8.0]-[24] indices, extending between 3.4-8.7 mag. We also show the reddening vector for 10 mag of visual extinction, and the likely vectors for increasing ionized, $H_2$ and 24 μm band emission. The BB dust continuum is indicated in the lower part of the graph, where individual ticks are separated by 10 K, whilst combined trends for PAH + dust emission are located to the right-hand side (dashed curves), where the trends correspond to differing values of $\Lambda$ (see text for details).

**Figure 6**

As for Fig. 5, but for the [5.8]-[8.0]/[3.6]-[4.5] colour plane. In this case, the PAH + dust emission loci extend throughout the colour plane.

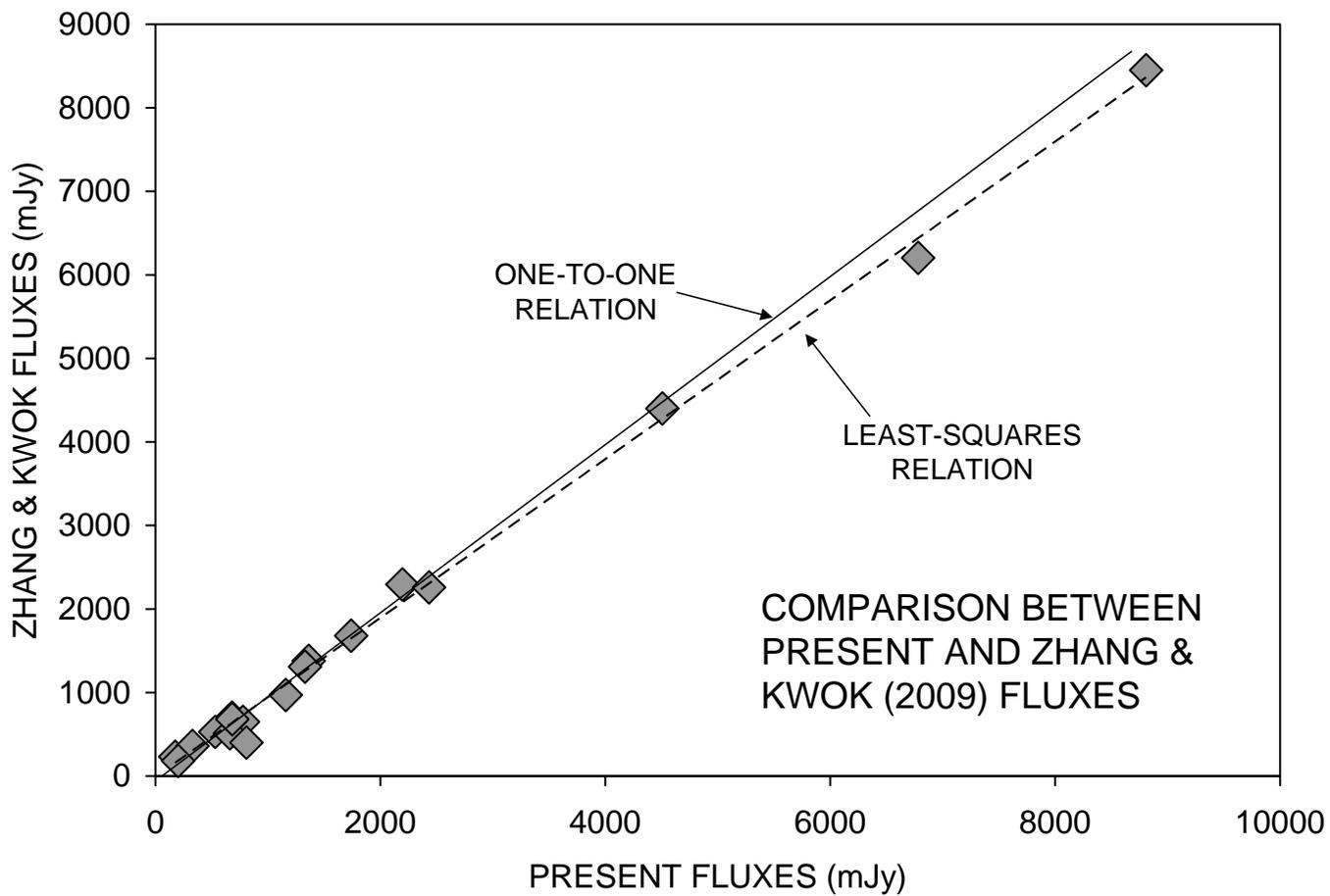
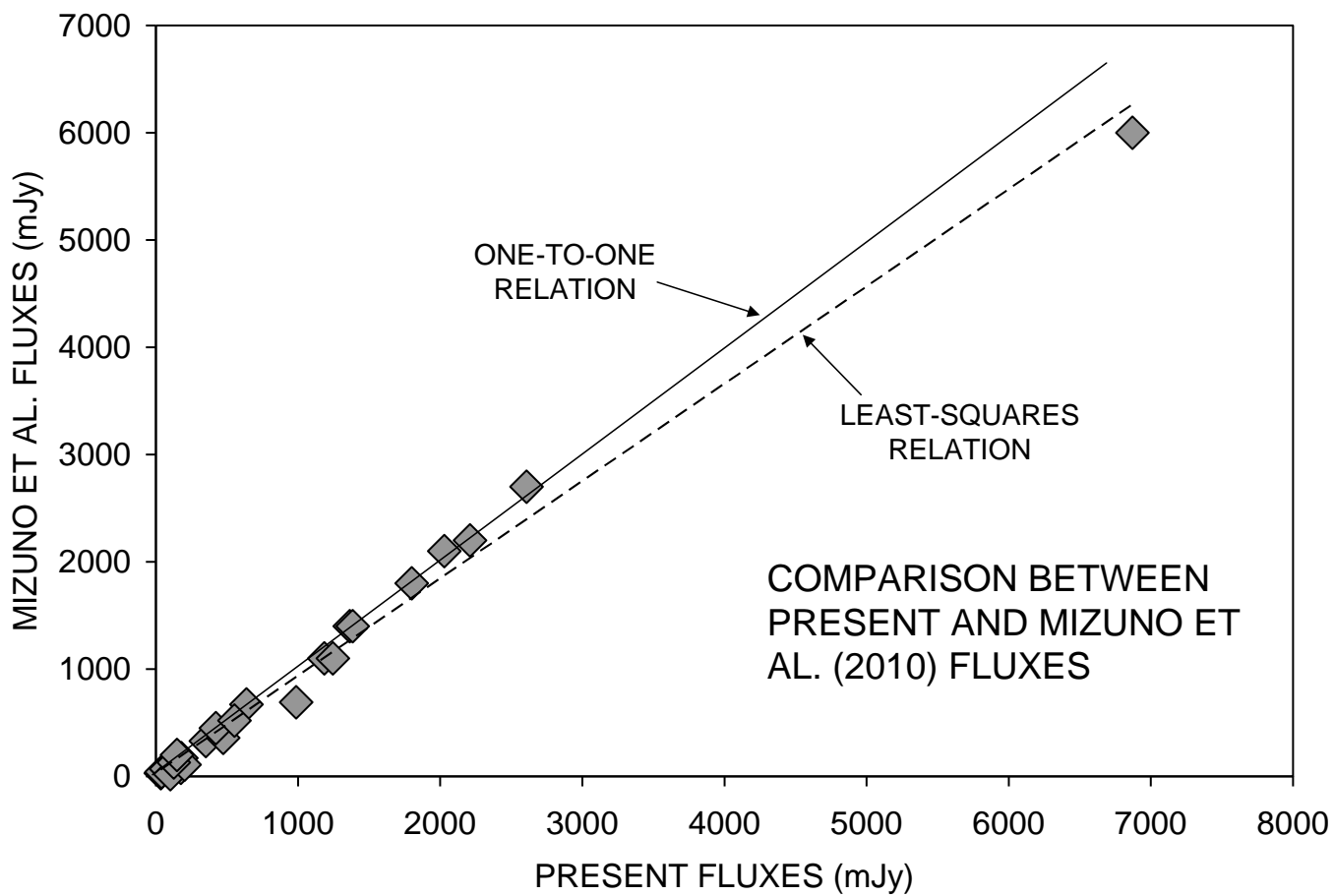

FIGURE 1

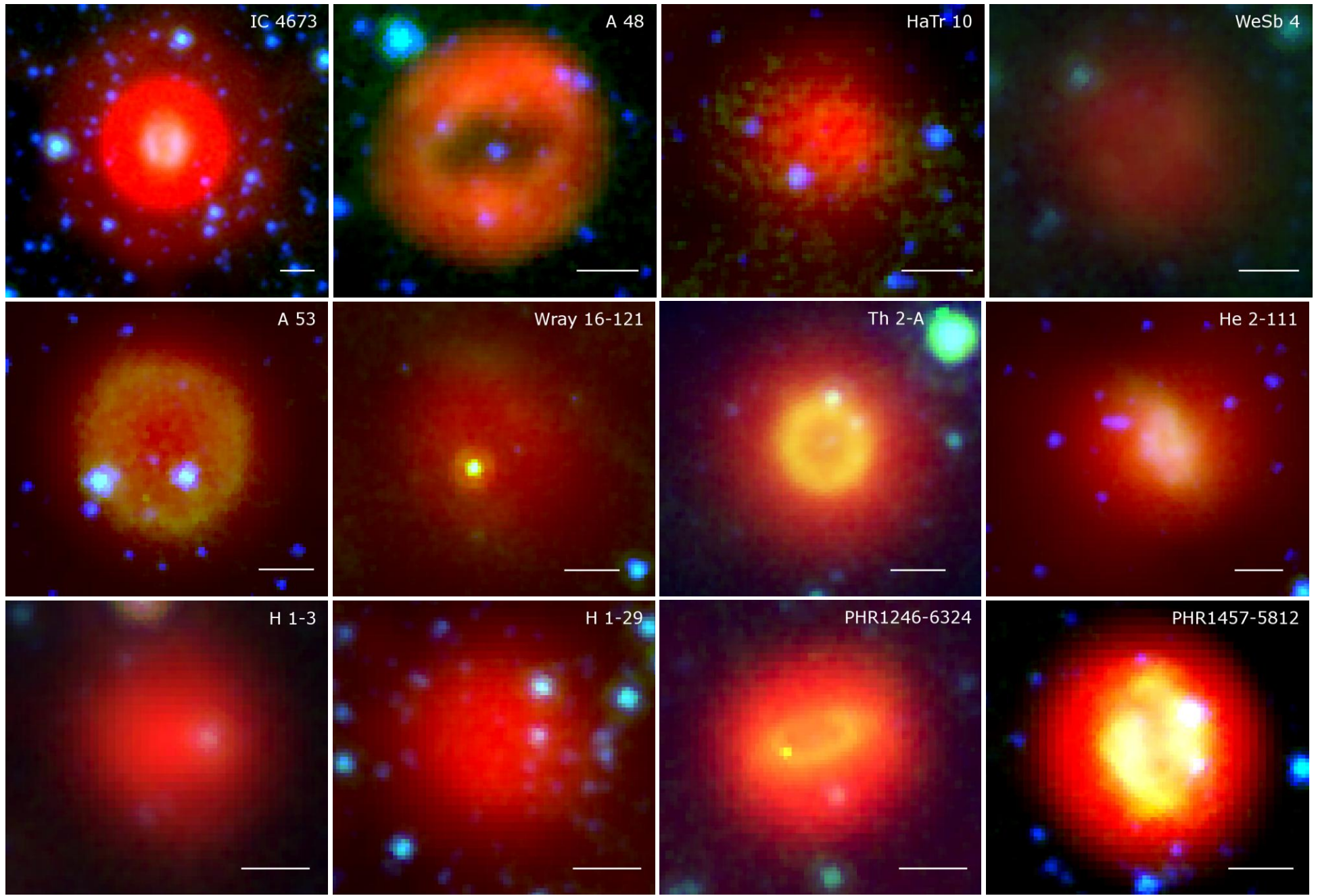

FIGURE 2

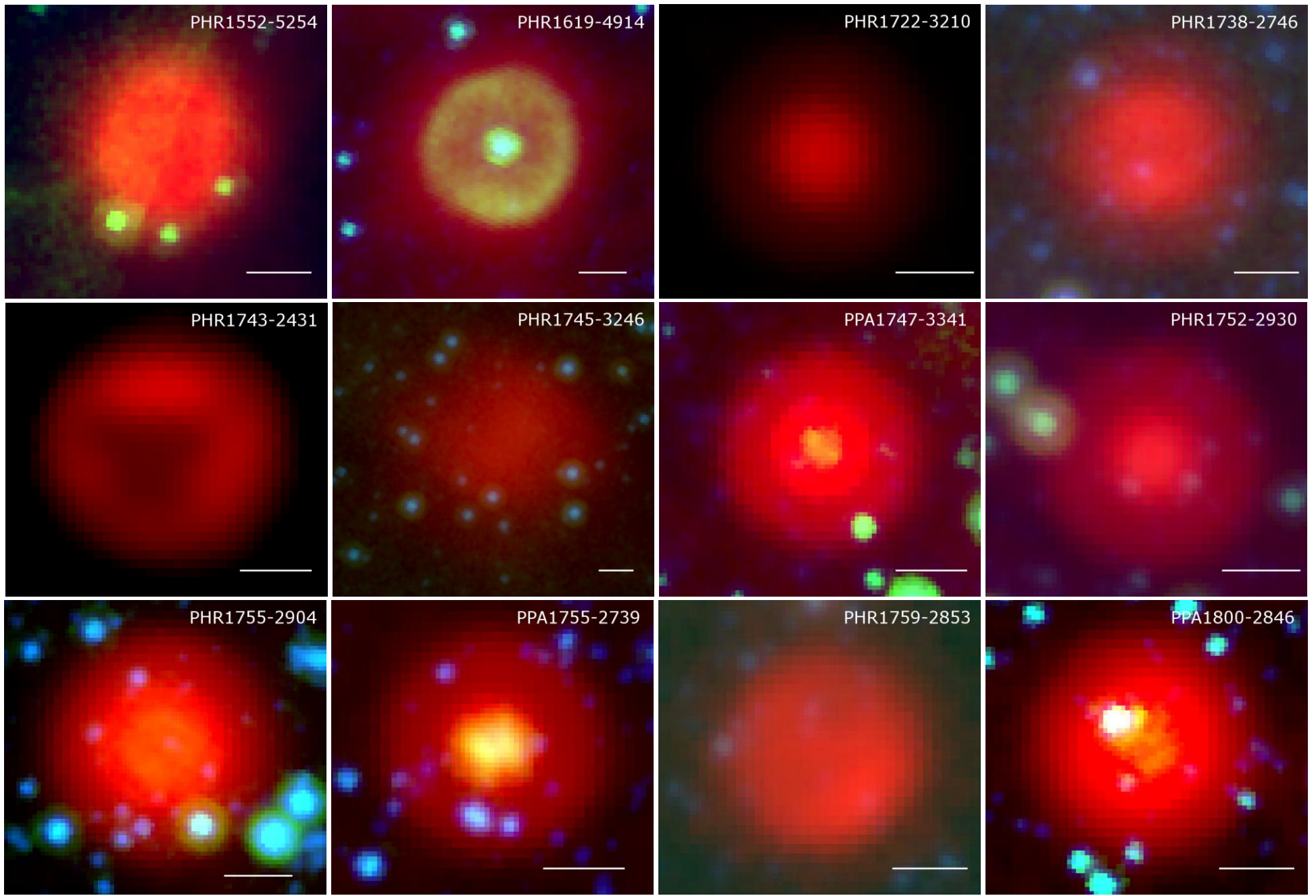

FIGURE 2 (CONT.)

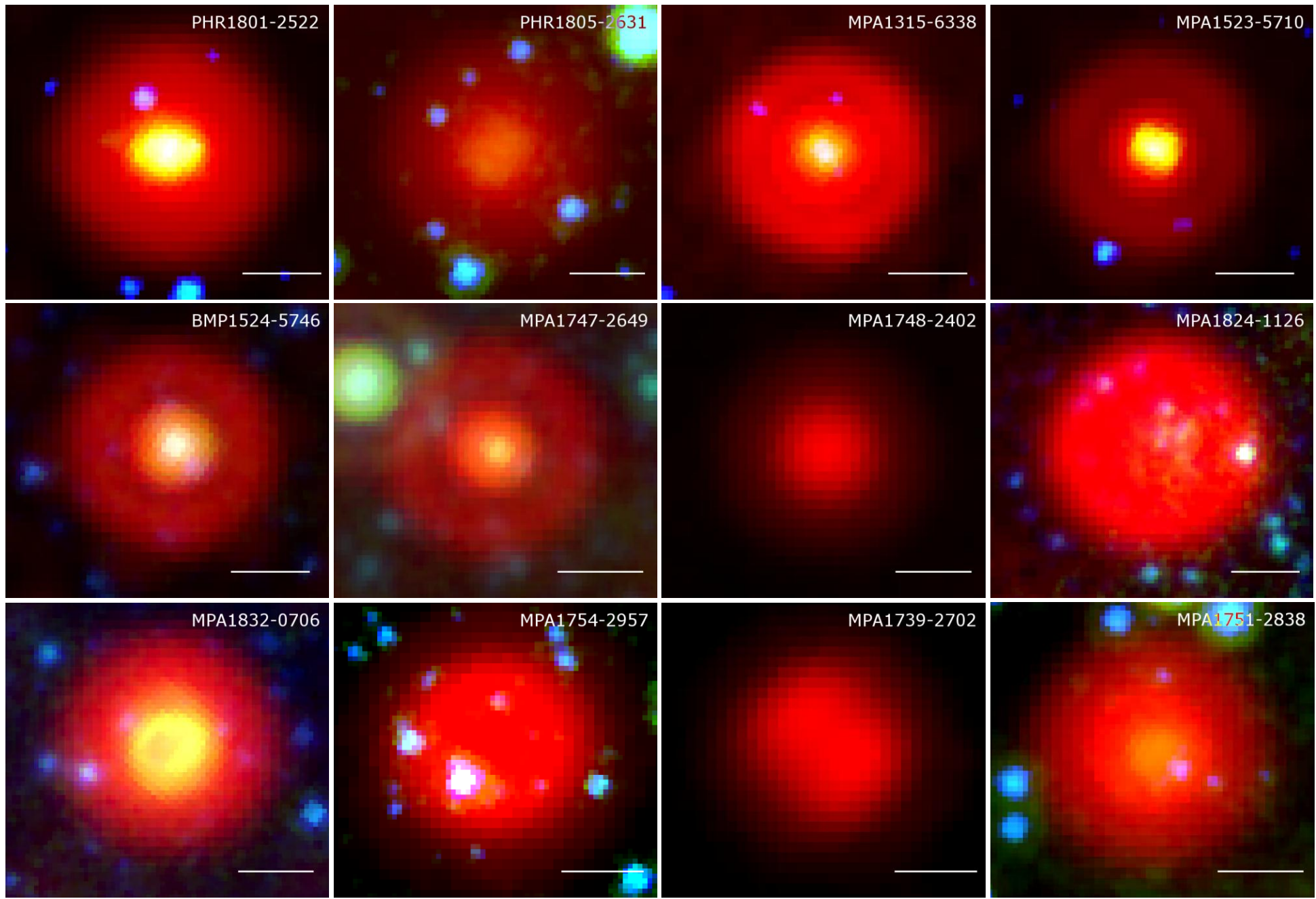

FIGURE 2 (CONT.)

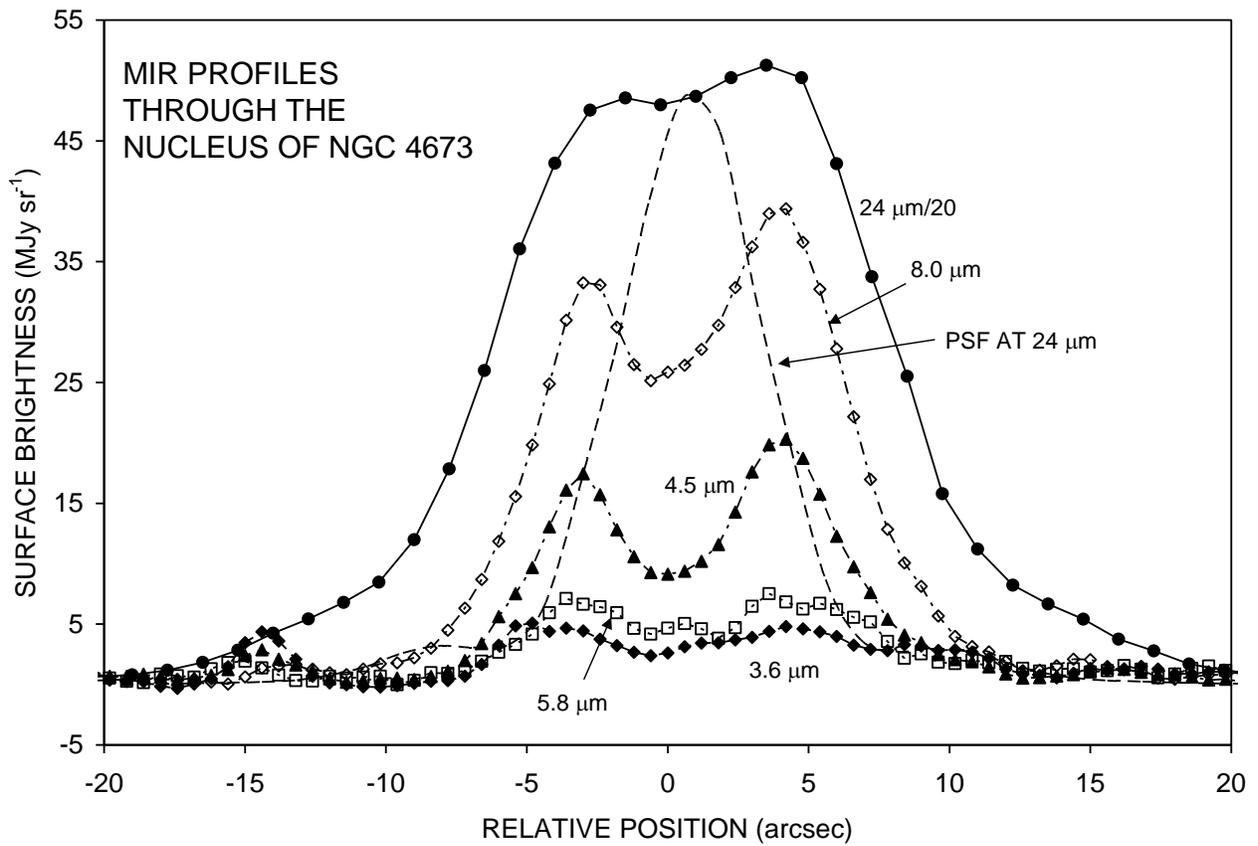

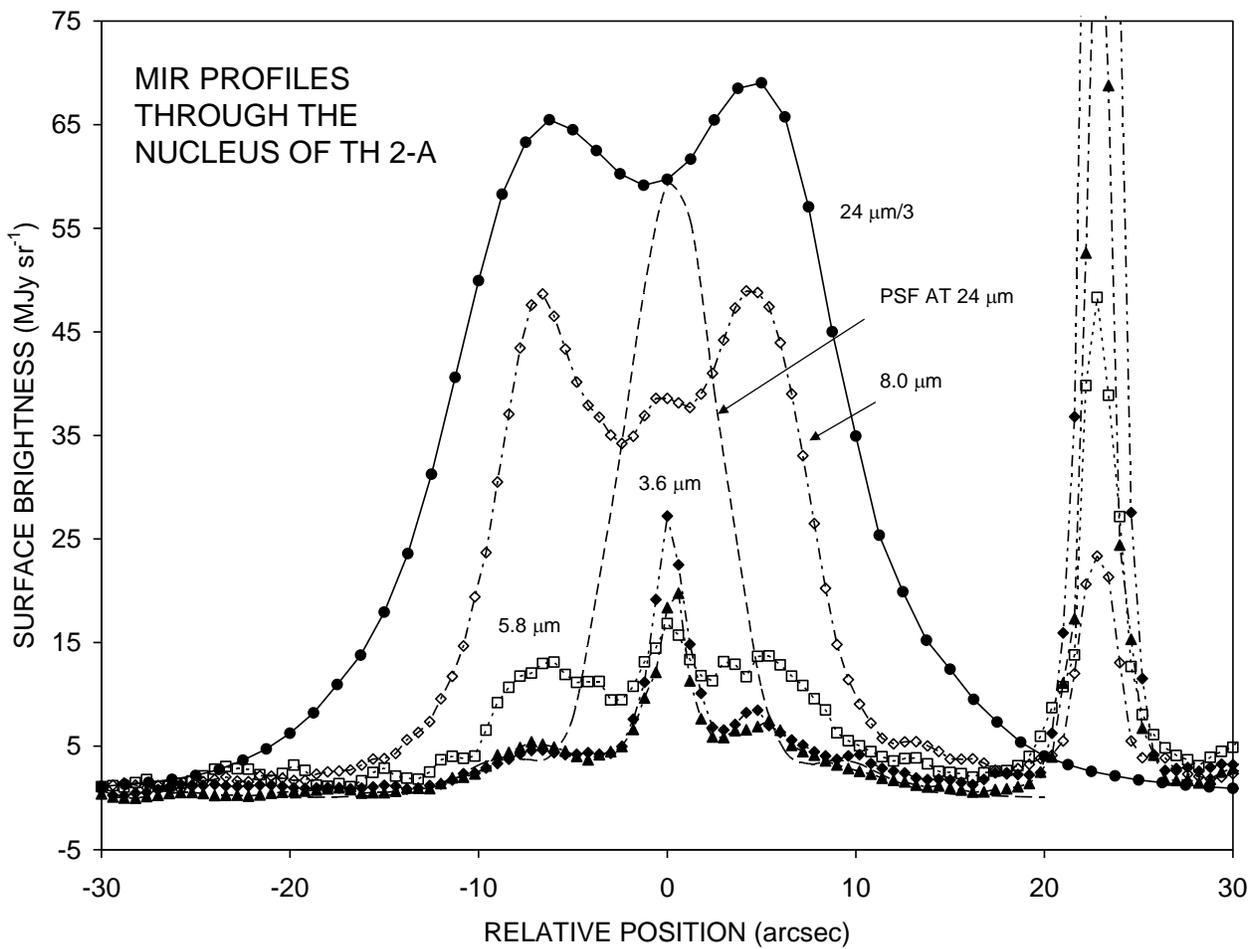

FIGURE 3

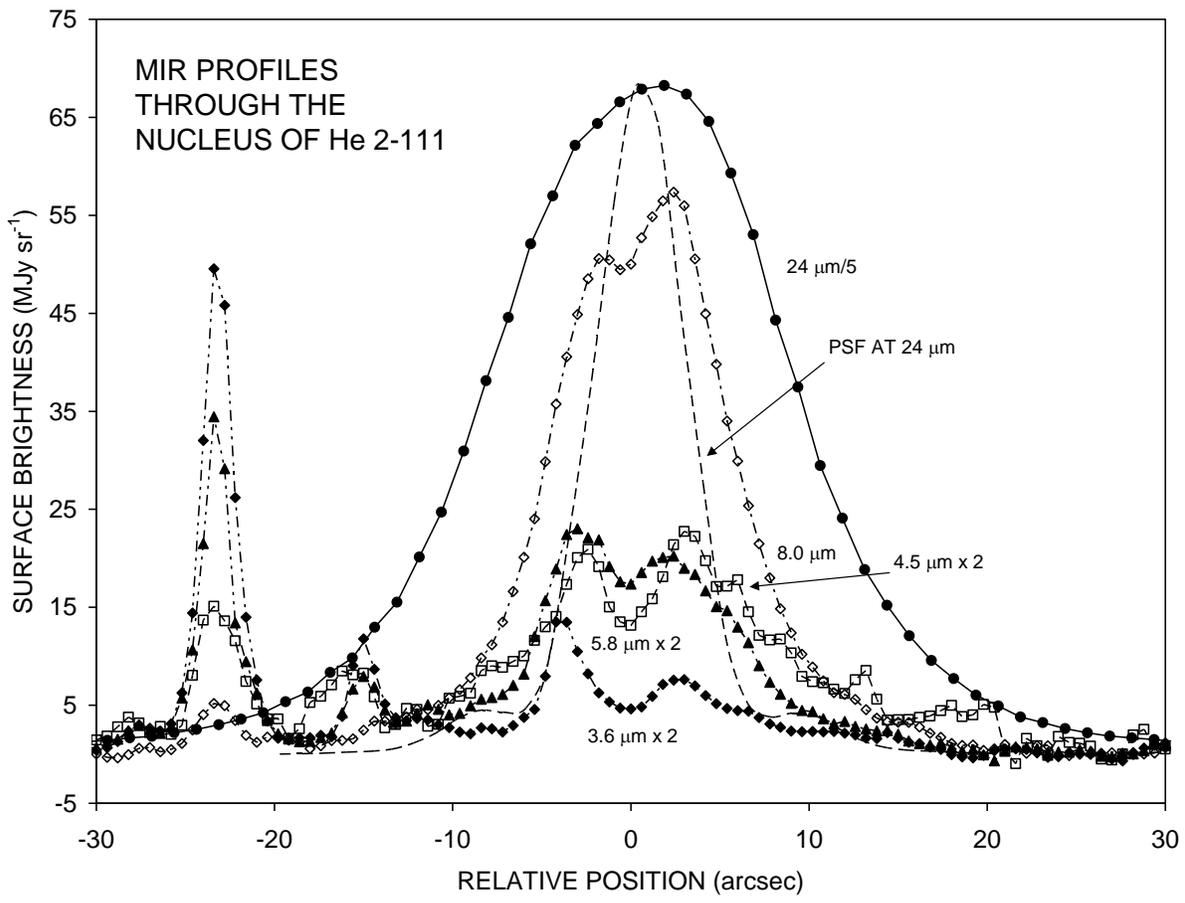

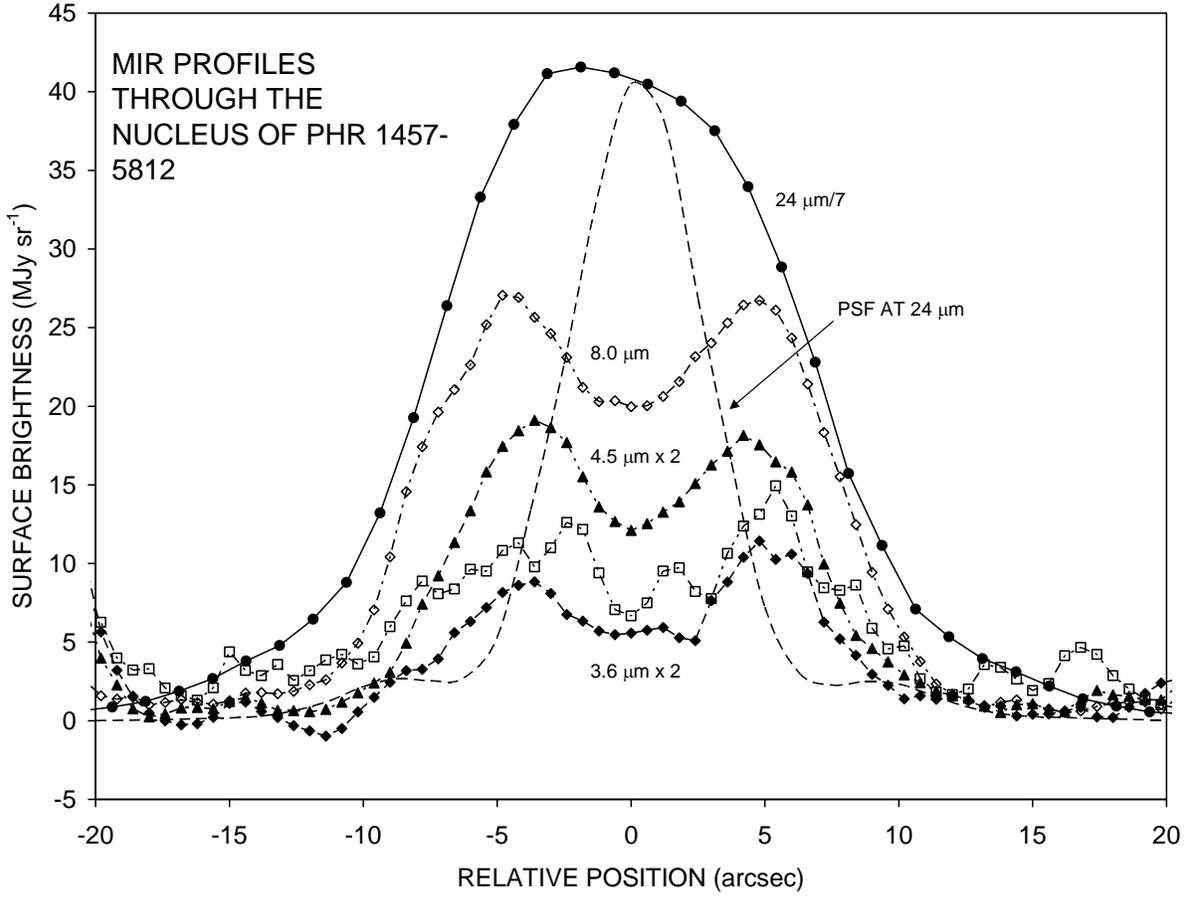

FIGURE 3 (CONT.)

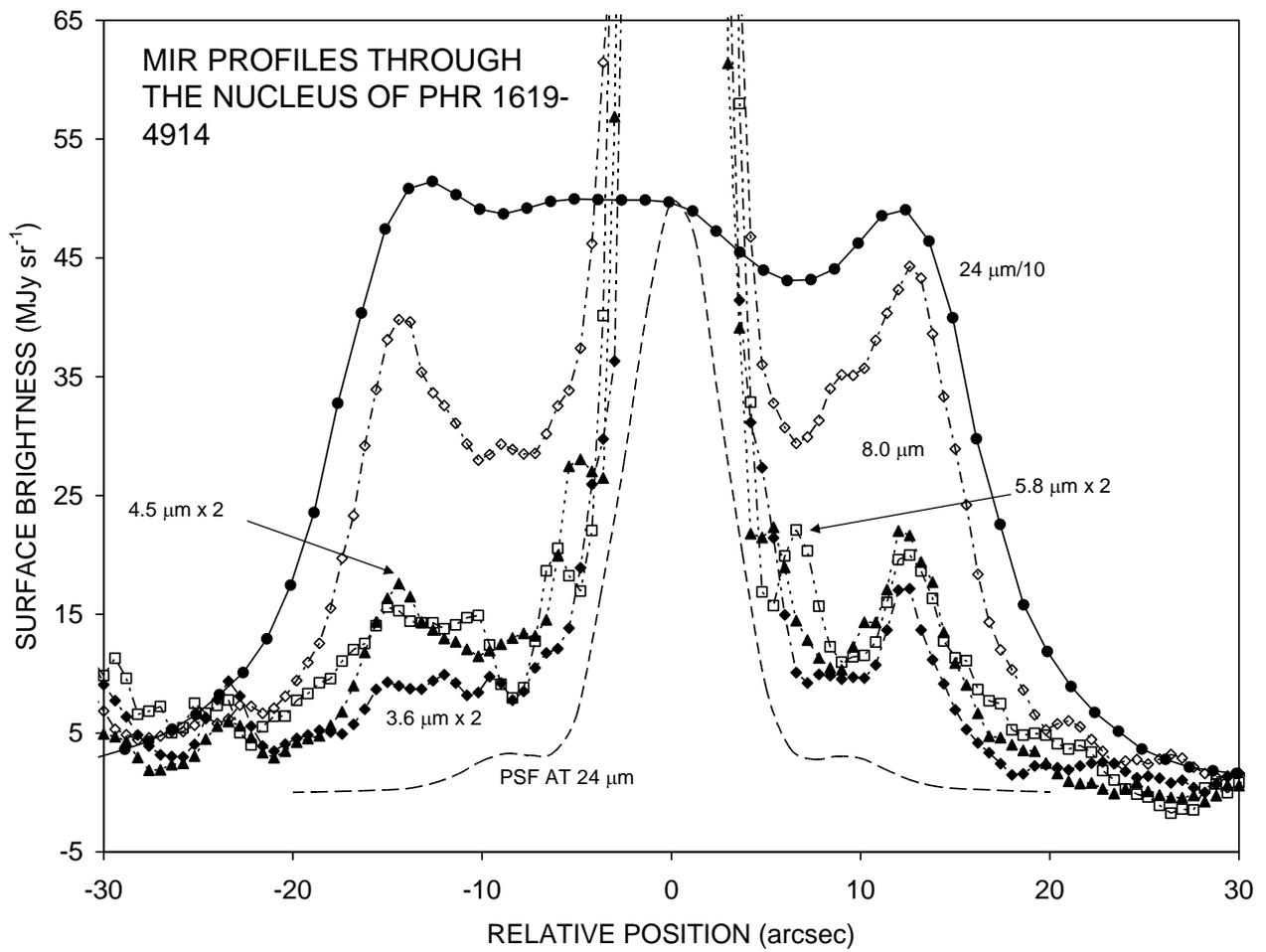
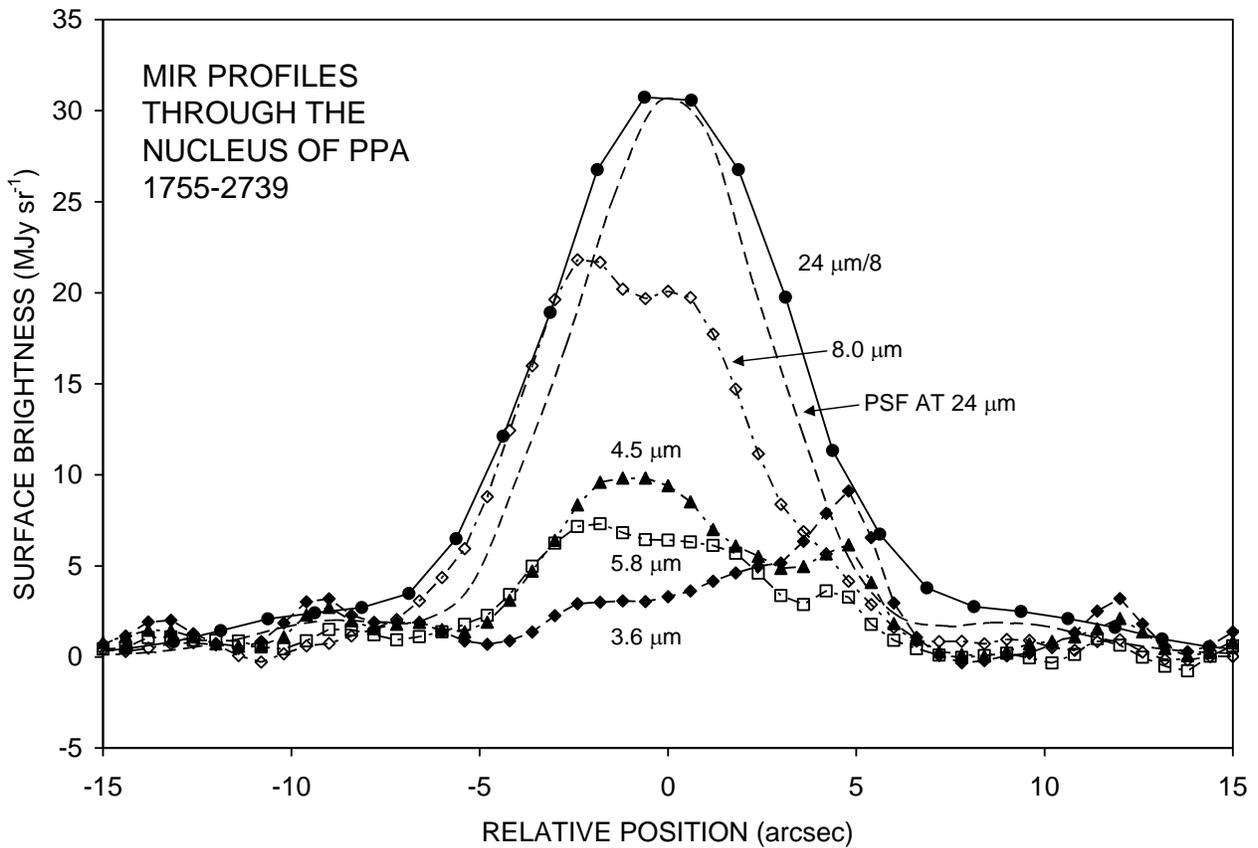

FIGURE 3 (CONT.)

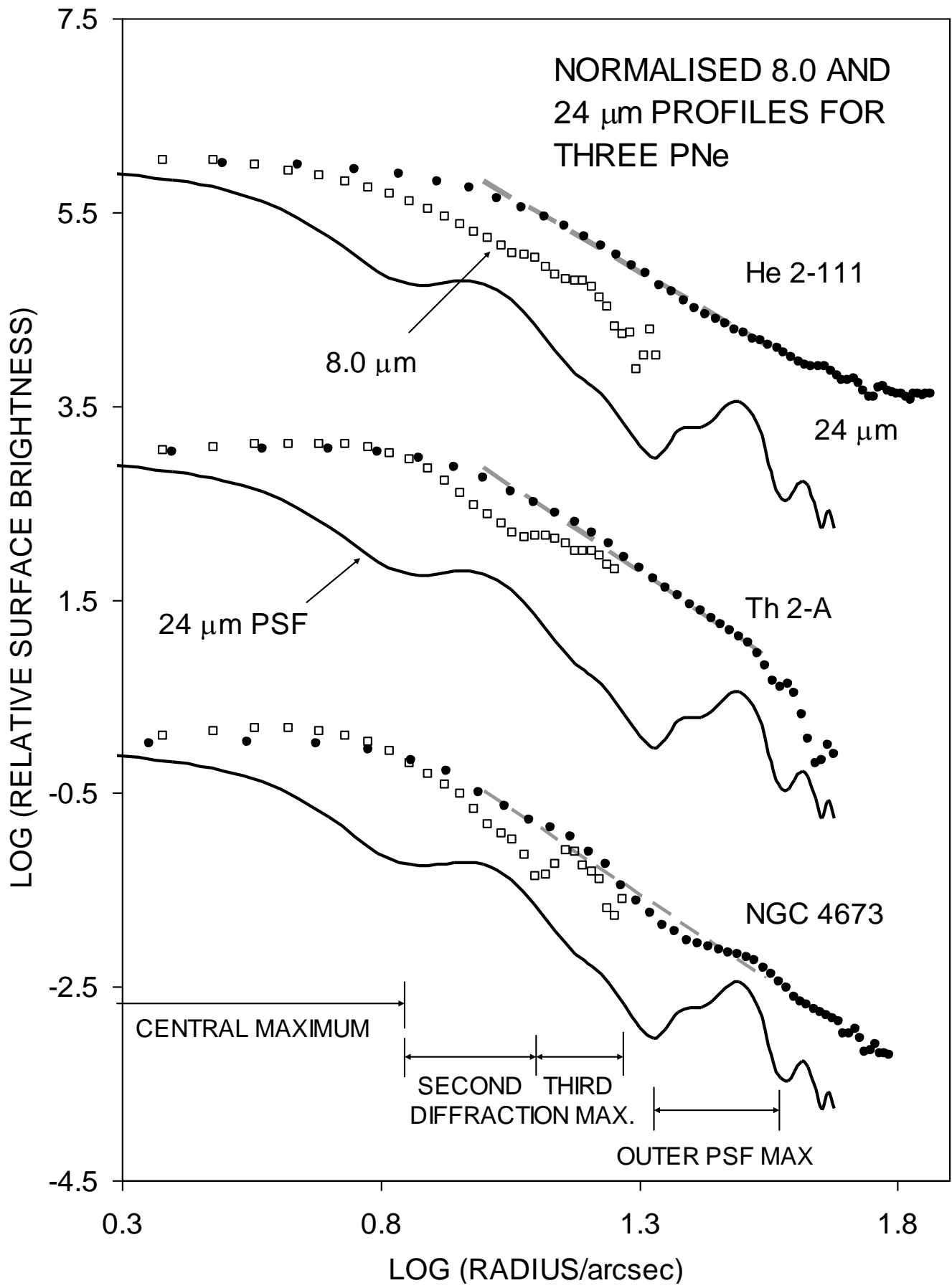

FIGURE 4

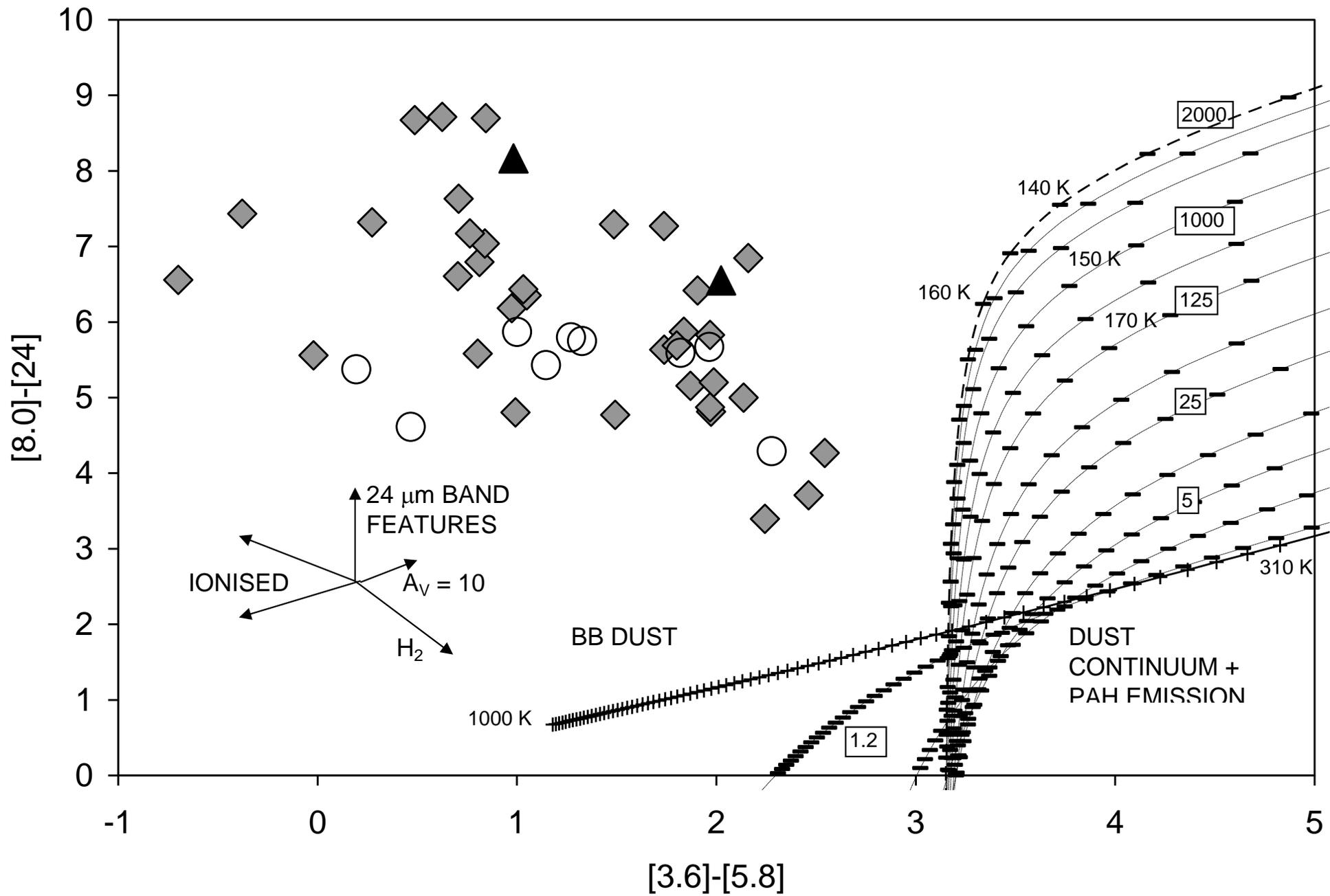

FIGURE 5

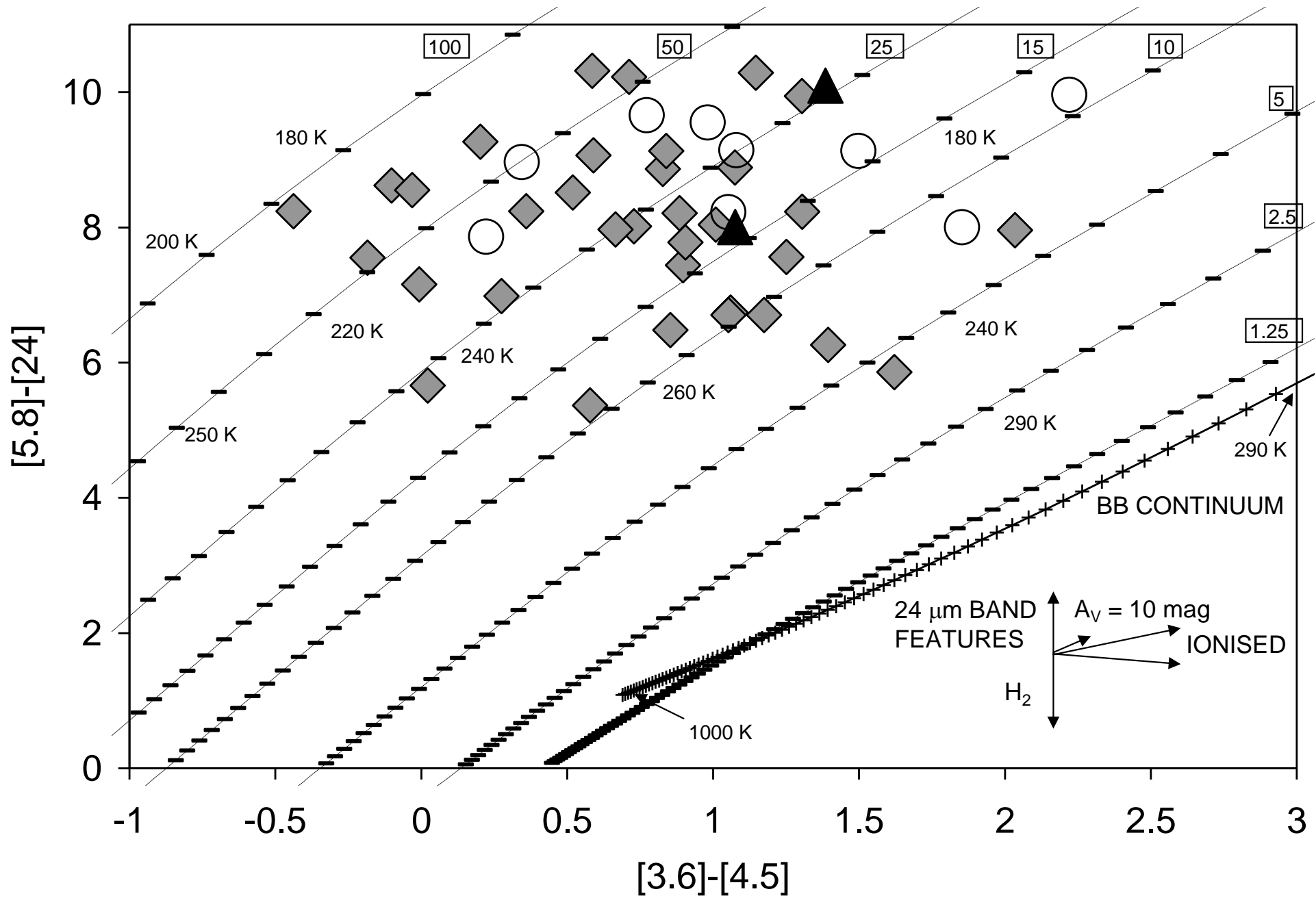

FIGURE 6